\documentclass[a4paper,11pt]{article}
\usepackage{jcappub} 
\usepackage{orcidlink}
\usepackage{multirow}
\usepackage{color, xcolor}

\newcommand{\disperse}{{\texttt{DisPeSE}}}
\newcommand{\rockstar}{{\texttt{Rockstar}}}

\newcommand{\mpch}{h^{-1} {\rm Mpc}}

\newcommand{\gpch}{h^{-1} {\rm Gpc}}

\newcommand{\Msun}{M_{\odot}}

\newcommand{\fsizefour}{3.7cm}

\title{\boldmath Artifacts in Halo Shapes: Imprints of the Initial Condition}

\author[~\orcidlink{0000-0002-9359-7170}\,a,c,d,1]{Yu Yu\note{Corresponding author.}}
\author[~\orcidlink{0000-0002-2183-9863}\,b,a,c,d]{, Zhao Chen}

\affiliation[a]{\textit{Department of Astronomy, School of Physics and Astronomy, Shanghai Jiao Tong University, 800 Dongchuan Road, Shanghai 200240, China}}
\affiliation[b]{\textit{Tsung-Dao Lee Institute $\&$ School of Physics and Astronomy, Shanghai Jiao Tong University, Shanghai 200240, China}}
\affiliation[c]{\textit{State Key Laboratory of Dark Matter Physics, School of Physics and Astronomy, Shanghai Jiao Tong University, 800 Dongchuan Road, Shanghai 200240, China}}
\affiliation[d]{\textit{Key Laboratory for Particle Astrophysics and Cosmology (MOE),
\& Shanghai Key Laboratory for Particle Physics and Cosmology, Shanghai Jiao Tong University, Shanghai, 200240, China}}

\emailAdd{yuyu22@sjtu.edu.cn}
\emailAdd{chyiru@sjtu.edu.cn}

\abstract{
Grid type pre-initial conditions are commonly used to initialize particle positions in cosmological simulations.
While these conditions are known to produce noticeable numerical artifacts in void regions, their impact on halo properties has generally been assumed to be negligible.
In this work, we employ multiple simulations to demonstrate that grid initialization induces statistically significant artifacts in halo shapes, despite the modest absolute amplitude ($\sim 1\%$) making them unimportant for most cosmological studies.
We identify a redshift-dependent artificial alignment pattern: at low redshifts ($z<2$), halo shapes preferentially orient away from the simulation box's Cartesian axes, whereas their constituent particles initially exhibit alignment with these axes. We propose a mathematical hypothesis to explain this flipping behavior.
}

\begin{document}
\maketitle
\flushbottom

\section{Introduction}
\label{sec:intro}

Cosmological simulations have become indispensable tools in modern cosmology and large-scale structure studies. 
They enable us to discriminate competing cosmological models, establish the structure formation paradigm, investigate halo properties in detail, and model complex nonlinear evolution processes (e.g., \cite{White83,White87, Bond96,Navarro97,Bertschinger98,Springel05}, see \cite{Angulo22} for a recent review on simulations).
In the era of precision cosmology, high-fidelity mock catalogs constructed from simulations play a dual role: they quantify the observational systematics \cite{Smith24,Blake25,Fernandez-Garcia25}, and model the astrophysical contamination.

A key astrophysical systematic in weak lensing studies is galaxy intrinsic alignment~\cite{Croft00,Heavens00}.
The galaxy shape prefers to trace the shape of its host halos/subhalos, and even the morphology of large-scale structures, and thus intrinsically correlates with each other.
While intrinsic alignment was traditionally treated as a contaminant, it has recently emerged as an independent cosmological probe (e.g., \cite{Xu23}). 
Similarly, the cosmological information encoded in cosmic web structures, particularly voids and filaments, has gained attention (e.g., \cite{Lavaux12, Wu25}).
However, theoretical predictions for these quasi-linear structures rely heavily on simulations, making it interesting to identify numerical artifacts in morphological probes (e.g., spurious alignments).

One source of numerical artifacts in cosmological simulations is the pre-initial conditions (pre-IC, or particle loads), i.e. the particle configuration at $z=\infty$.
The two most widely used schemes are (i) the anisotropic grid-based initial loads (e.g., \cite{Efstathiou85}), and (ii) the isotropic glass-like distribution \cite{White94}, generated via anti-gravity simulations.
While placing particles on a Cartesian grid is computationally straightforward, this approach introduces well known artifacts in void regions even in the late time~\cite{Baugh95}. 
Near void centers where particle displacements remain small, the initial grid pattern remains visibly imprinted in the particle distribution. 
Although these artifacts are easily identifiable by eye, their statistical impact on void profiles has received limited quantitative study.

The impact of initial condition artifacts has been extensively studied in hot dark matter simulations (e.g., \cite{Centrella88,Gotz03,Wang07}).
In these scenarios, the absence of small-scale fluctuations in the initial conditions preserves the initial particle pattern in quasi-linear structures such as filaments.
This leads to artificial filament fragmentation and spurious substructures, even when non-grid particle loading schemes are employed~\cite{Wang07}. 

In contrast, cold dark matter simulations have received less attention regarding these artifacts,
either because nonlinear evolution was presumed to erase them, or because their effects were deemed negligible.
Study comparing grid- and glass-type initial condition shows differences of less than $1\%$  in matter power spectrum and halo mass function~\cite{LHuillier14}.
However, separate universe simulations reveal an enhanced anisotropic tidal response at $z\ge 9$ when grid-type loads are used~\cite{Masaki21}.
They suggested that isotropic statistical quantities such as the matter power spectrum is not a good indicator of these artifacts,
while the tidal response function at high redshifts is a good example where an appropriate choice of pre-ICs is crucial.

Beyond the conventional grid and glass initial loads, alternative schemes such as Q-set~\cite{Hansen07}, Bravais lattices~\cite{Joyce09} and CCVT (capacity constrained Voronoi tessellation)~\cite{Liao18} have been proposed.
While these different initialization methods do produce measurable differences in simulation outputs (e.g., in halo positions and properties~\cite{Zhang21}), the field has yet to establish consensus on an optimal approach.
Particularly relevant to our study, the common expectation that halo particles completely lose memory of their initial configuration during virialization has led to the theoretical presumption that halo shapes should be insensitive to these numerical effects. 
However, as we demonstrate in this work, this assumption warrants careful re-examination.

In this work we attempt to identify the numerical artifacts in halo shapes.
Our analysis reveals statistically significant artifacts when using grid-type initial conditions, while simulations with glass-type initial loads show no detectable effects.
This paper is organized as follows.
Section~\ref{sec:method} presents our methodology for identifying numerical artifacts in both halo shapes and filament orientations.
In section~\ref{sec:results} we report the identified numerical artifacts, and their underlying causes.
Section~\ref{sec:conclusion} summarizes our findings and discusses directions for future research.

\section{Methodology}
\label{sec:method}

\subsection{Simulations}
\label{sec:simu}

In this work we investigate potential artifacts in simulated halo shapes using various simulations.

CosmicGrowth is a high-resolution simulation suite spanning multiple cosmological models and configurations.
Specifically, we analyze the ``WMAP\_3072\_600'' run
(see table 4 in ref.~\cite{Jing19}),
which tracks the evolution of $3072^3$ particles in a cubic volume with a side length of $600\,\mpch$.
The simulation adopts the following cosmological parameters: matter density today $\Omega_m=0.268$, cosmological constant density $\Omega_\Lambda=0.732$, dimensionless Hubble parameter $h=0.71$, primordial power spectrum tilt $n_s=0.968$ and linear power spectrum amplitude $\sigma_8=0.83$.
Each particle has a mass of $5.54\times 10^8 \Msun/h$. The initial conditions are generated at $z=144$ using 1st-order Lagrangian Perturbation Theory (1LPT), with grid pre-IC particle loads.

ELUCID\footnote{\url{https://gax.sjtu.edu.cn/data/ELUCID.html}}~\cite{wang16} is a constrained simulation based on reconstructed initial condition from the observed SDSS volume. 
It evolves $3072^3$ particles in a $500\,\mpch$ box, adopting WMAP cosmology with $\Omega_m=0.258$, $\Omega_\Lambda=0.742$, $h=0.72$, $n_s=0.963$ and $\sigma_8=0.796$.
The particle mass is $3.09\times 10^8 \Msun/h$,
and initial conditions are generated at $z=100$ using 1LPT with grid-type pre-IC.

Uchuu1000Pl18\footnote{\url{https://skiesanduniverses.org/Simulations/Uchuu/}}, part of the Uchuu simulation suite~\cite{Ishiyama21}, models $6400^3$ particles in a $1\,\gpch$ volume with Planck cosmology:
$\Omega_m=0.3089$, $\Omega_\Lambda=0.6911$, $h=0.6774$, $n_s=0.9667$ and $\sigma_8=0.8159$.
Particle mass is $3.29\times 10^8\Msun/h$,
and initial conditions are set at $z=127$ using 2LPT.

MDPL2\footnote{\url{https://www.cosmosim.org/metadata/mdpl2/}} (MultiDark Planck 2)~\cite{Klypin16} evolves $3840^3$ particles in a $1\,\gpch$ box, with
$\Omega_m=0.307115$, $\Omega_\Lambda=0.692885$, $h=0.6777$, $n_s=0.96$ and $\sigma_8=0.8228$.
Particle mass is $1.51\times 10^9 \Msun/h$,
and the 1LPT initial conditions are applied at $z=120$ with grid pre-loading.

Kun Universe\footnote{\url{https://kunsimulation.readthedocs.io/}} comprises 129 high-resolution simulations for cosmological emulation~\cite{Chen25,Chen25a,Zhou25}, 
and each simulation evolves $3072^3$ particles in a $1\,\gpch$ box using different cosmologies.
We analyze its fiducial run with cosmological parameters $\Omega_m=0.3111$, $\Omega_\Lambda=0.6889$, $h=0.6766$, $n_s=0.9665$, $\sigma_8=0.81$, the dark energy equation-of-state parameters $w_0=-1$, $w_a=0$, and the neutrino mass $\sum m_\nu=0.06$.
Particle mass of the fiducial run is $2.96\times 10^9 \Msun/h$, and the initial redshift is $z=127$.
Differently from all the simulations above, the pre-initial particle loading in Kun is glass-type.

A summary of these simulations is provided in table~\ref{tab:simuinfo}.

\begin{table*}
\centering
\caption{
The simulations used in this work.
\label{tab:simuinfo}
}
\begin{tabular}{c|c|c|c|c|c}
\hline
Simulation & Halo Finder & Pre-initial & Box size & Code & IC code\\
\hline
CosmicGrowth & FoF & Grid & $600$ & Jing (P3M)& Jing\\
ELUCID & FoF & Grid & $500$ & Gadget-2 (TPM) & NGenIC \\
Uchuu1000Pl18 & Rockstar & Grid & $1000$ & GreeM (TPM) & 2LPTic \\
MDPL2 & Rockstar & Grid & $1000$ & Gadget-2 (TPM) & Ginnungagap \\
Kun fiducial & FoF & Glass & $1000$ & Gadget-4 (TPM) & Gadget-4\\
\hline
\end{tabular}
\end{table*}

\subsection{Halo Finding and Shape Definition}
\label{sec:shape}

The friends-of-friends (FoF) algorithm is a widely used method for identifying halos in cosmological simulations.
It links particles separated by distances smaller than a specified threshold, defining halos as compact, gravitationally bound objects.
In this work, we analyze halos from the  CosmicGrowth, ELUCID, Kun fiducial simulations, each of which employs its own FoF implementation with a common linking length parameter $b=0.2$.
While subtle differences exist between these FoF algorithms (e.g., in the treatment of unbound particles), such variations are negligible for our purposes, as we do not compare halos across simulations or halo-finder implementations.

For each FoF halo, we compute the inertia tensor from the particle positions,
\begin{equation}
I_{ij}=\frac{1}{N}\sum x_i x_j\ ,
\end{equation}
where $x$ denotes the particle position relative to the halo center, $i$ and $j$ index Cartesian coordinates, and the summation runs over all $N$ particles in the halo.
Eigen-decomposition of this symmetric $3\times 3$ tensor yields three eigenvectors, which correspond to the halo's principal axes.
The major axis, associated with the largest eigenvalue, is of particular interest for studying numerical artifacts.
This axis is widely used to quantify halo alignment.
In appendix~\ref{sec:appendix_minor}, we also show part of the result for minor axes, which is expected to be more sensitive to numerical artifacts, should they exist.

For Uchuu1000Pl18 and MDPL2 simulation, we examine halos and subhalos identified by the \rockstar{} algorithm, a phase-space FoF halo finder that incorporates particle velocity information to mitigate the ``bridging effect'' (spurious linking of distinct clumps due to intervening particles).
\rockstar{} directly provides the major axes derived from inertia tensor decomposition.

\subsection{Filament-finding and Orientation Definition}
\label{sec:fila}

In the structure formation scenario,
halos are expected to migrate along filaments,
where mergers also frequently occur.
This framework is commonly invoked to explain the observed alignment between halo spin/angular momentum and filaments, as well as its dependence on halo mass (e.g., \cite{Bailin05,Codis15,Codis18,Wang25}).

To investigate the numerical origin of the halo shape artifacts, we identify filaments and their orientations using the CosmicGrowth simulation.
For this purpose, we employ the Discrete Persistent Structures Extractor (\disperse{})~\cite{Sousbie11,Sousbie11a}, a filament-finding algorithm based on discrete Morse theory and persistent homology.

\disperse{} formalizes the partitioning of space through the gradient flow of the density field, defining filaments as ascending 1-manifolds.
The method makes no assumptions about survey geometry or homogeneity and operates directly on discrete samples.
The identified filaments are wildly investigated to understand the structure formation (e.g., refs.~\cite{Wang24,Yang25}), environmental dependencies~\cite{Yu25}, alignment phenomena including angular momentum, galaxy shape and spin~\cite{Codis18, Tang25}, to find the missing baryons~\cite{Li25}, and the imprinted cosmological information (e.g., ref.~\cite{Wu25}).

In our implementation, we use halo catalogs (rather than dark matter particles) to trace filaments, significantly reducing computational costs.
We consider only FoF halos with $\ge 20$ member particles.
The resulting filaments from \disperse{} consist of linear segments connecting critical points.
We neglect the curving along filaments and define the filament orientation as the vector between the two critical points.

\subsection{Artifacts Visualization}
\label{sec:visual}

The nonlinear and complex gravitational evolution of dark matter particles tends to erase most numerical artifacts present in the initial condition.
To robustly detect any remaining subtle numerical effects, we employ the following statistic approach.

\emph{Vector Counting and Smoothing:} We count the number of vectors (halo major axes and the filament orientations) pointing toward each pixel in a HEALPix map\footnote{\url{https://healpix.sourceforge.io/}} (resolution parameter $N_\mathrm{side}=512$).
The resulting map is smoothed with a $10\deg$ Gaussian kernel to enhance any potential directional preferences.
This smoothing scale is optimized to best reveal these effects while maintaining sufficient resolution.

\emph{Vector Symmetry Handling:}
For each vector we account for both its positive and negative directions (i.e., treating them as equivalent).
This approach has two advantages: (i) It increases the signal-to-noise ratio by effectively doubling the sample size.
(ii) It ensures map symmetry, allowing us to present result in a half-sky mode using the HEALPix orthographic projection.

In addition to visual inspection of the smoothed maps, we quantify directional preferences by computing the probability excess of the cosine angle between each vector and the $x$, $y$, $z$-axis of the simulation box.
Again, both the positive and negative vector directions are accounted, and the results relative to all three axes are stacked to improve the statistical significance.

\section{Results}
\label{sec:results}

\subsection{Artifacts in Halo Shapes}
\label{sec:z-dependence}

\begin{figure*}[htbp]
\centering
\includegraphics[width=\fsizefour]{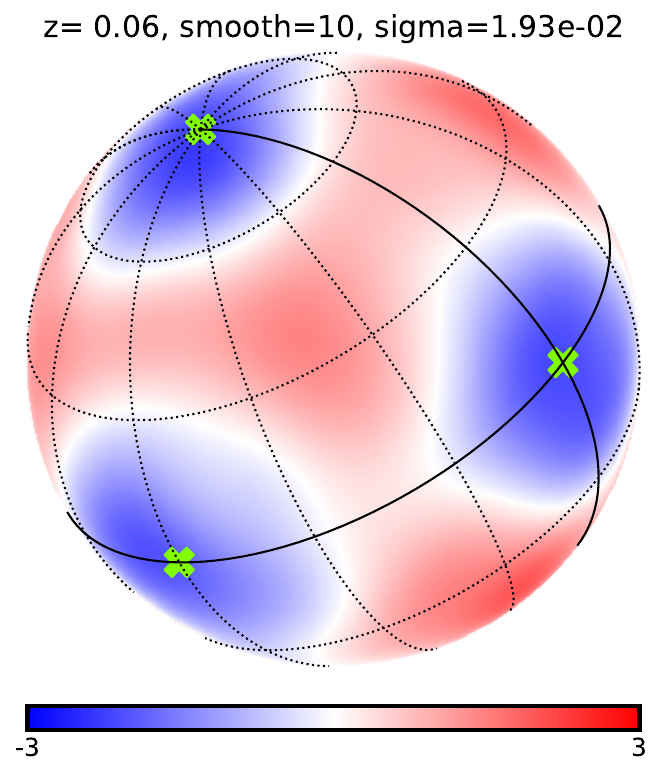}
\includegraphics[width=\fsizefour]{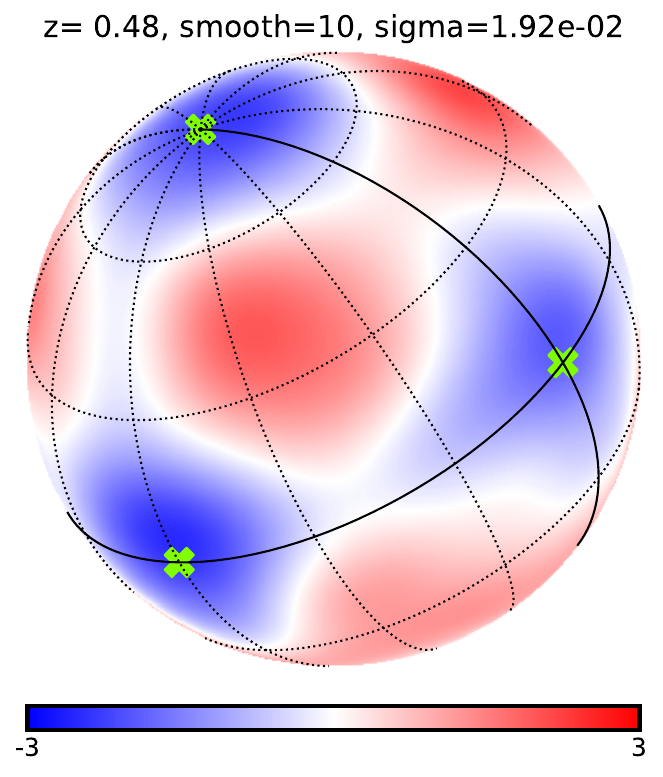}
\includegraphics[width=\fsizefour]{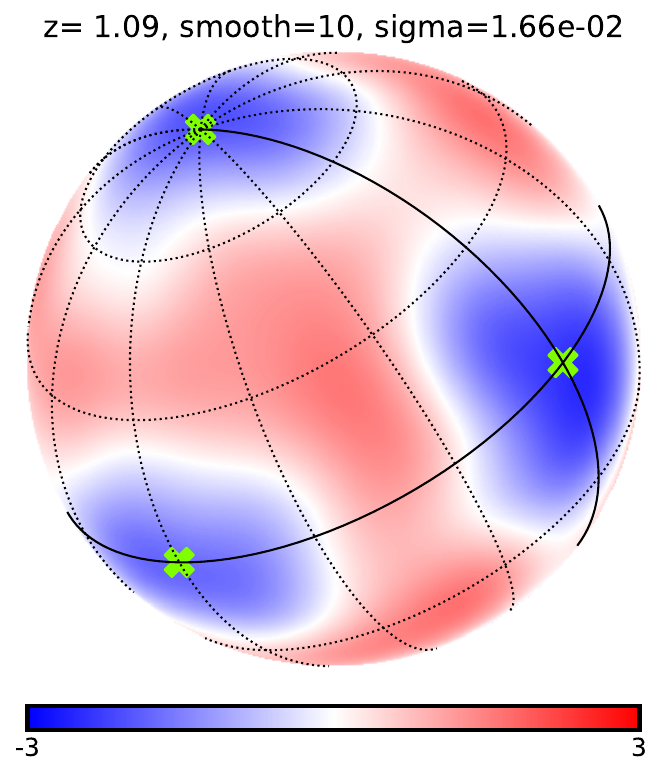}
\includegraphics[width=\fsizefour]{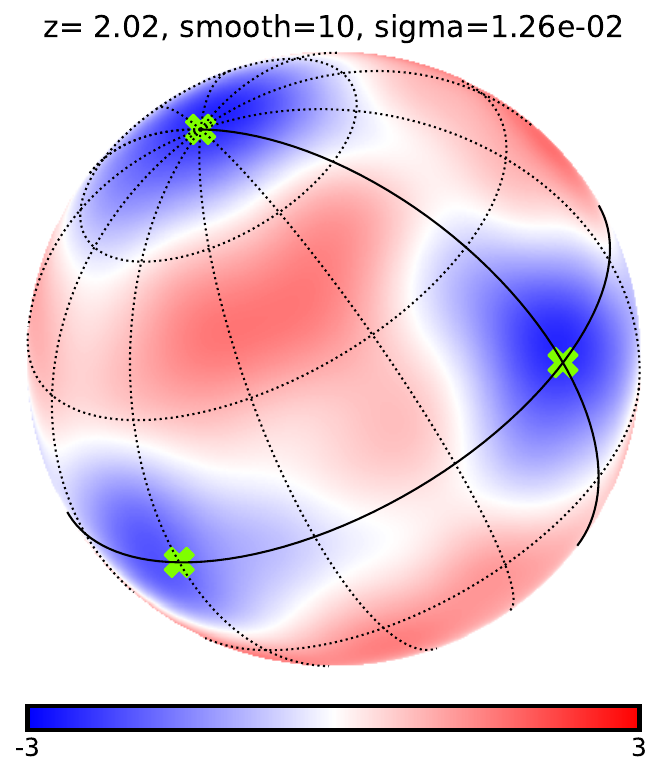}
\caption{
The halo shape artifacts in CosmicGrowth simulation.
Columns from left to right present the results for multiple redshifts from $z=0.06$ to $z=2.02$.
All the FoF halos with $\ge 200$ member particles are used.
The red/blue color represents the regions that halos prefer/avoid to point.
The green cross markers indicate the Cartesian axes of the simulation box.
The results show that halos slightly prefer \emph{not} to point at the three main Cartesian axes.
}
\label{fig:6610}
\end{figure*}

\begin{figure*}[htbp]
\centering
\includegraphics[width=\fsizefour]{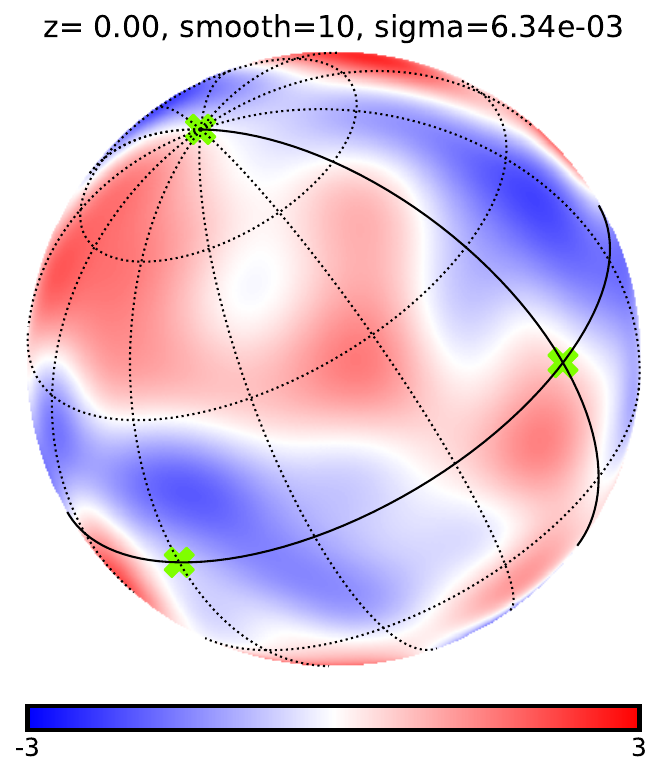}
\includegraphics[width=\fsizefour]{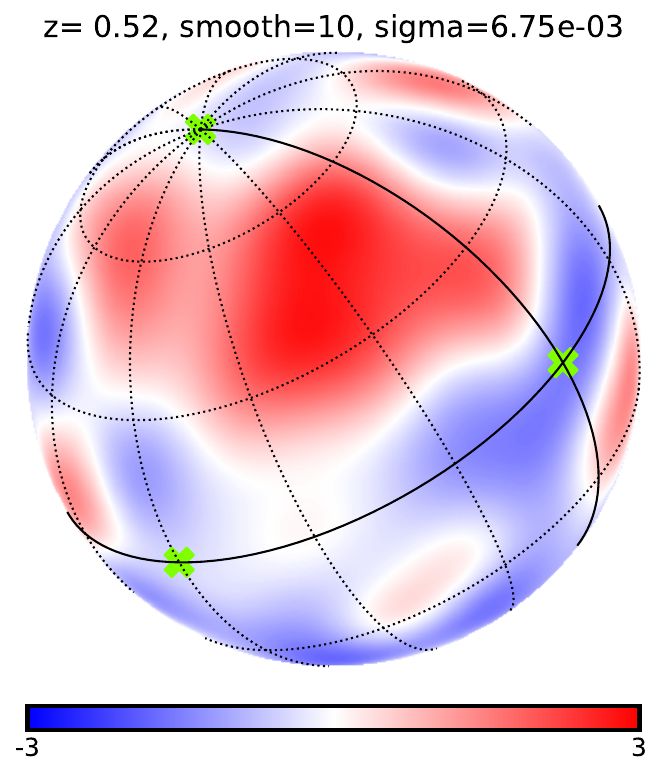}
\includegraphics[width=\fsizefour]{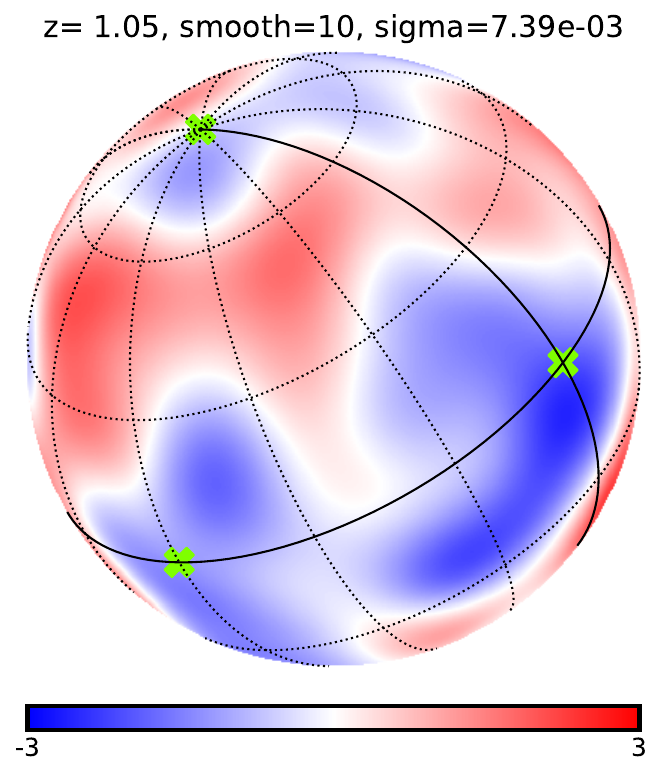}
\includegraphics[width=\fsizefour]{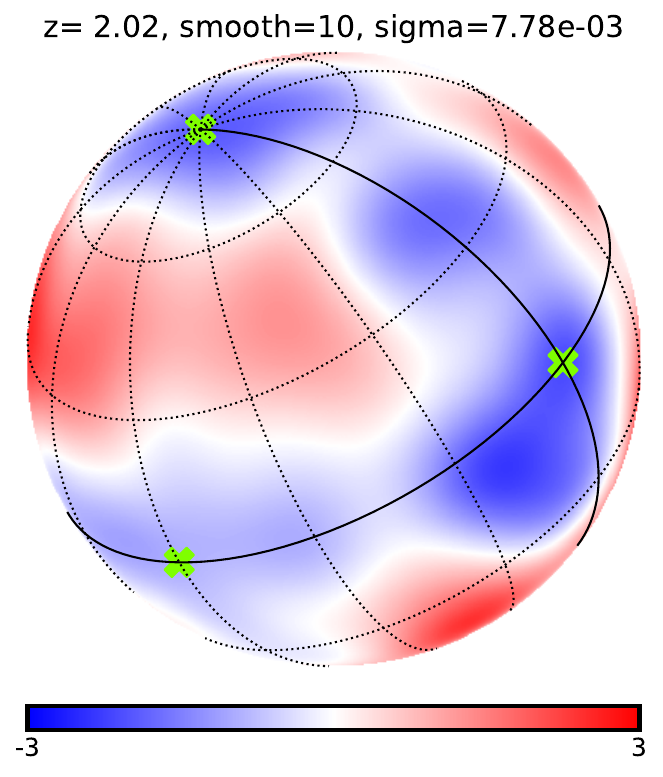}
\\
\includegraphics[width=\fsizefour]{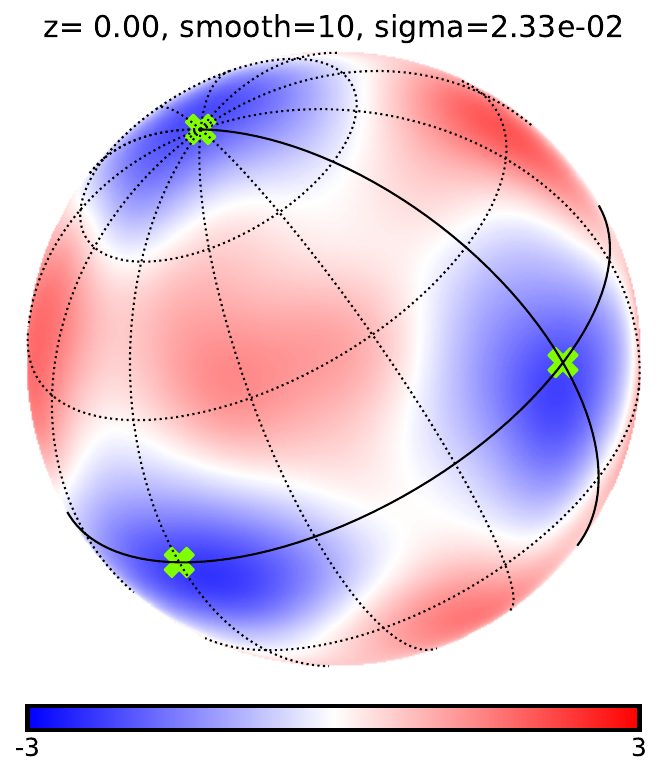}
\includegraphics[width=\fsizefour]{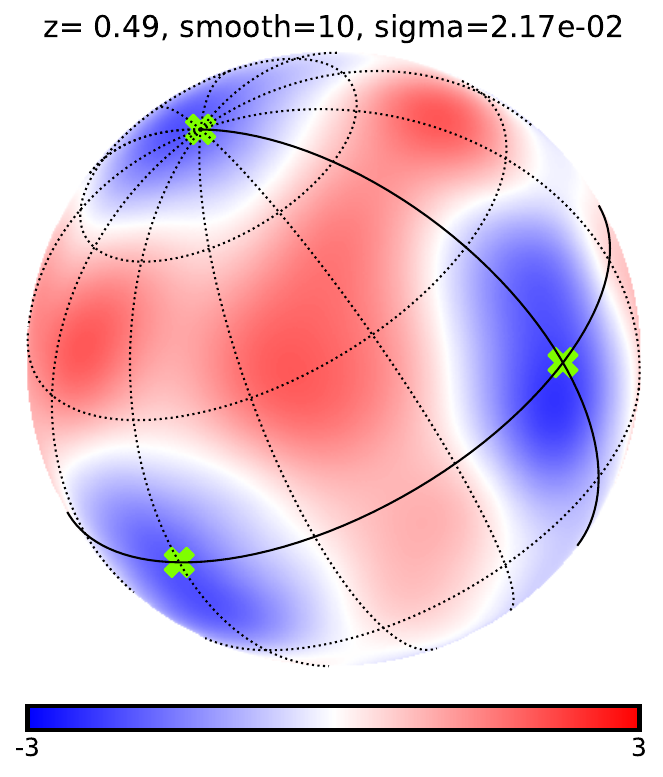}
\includegraphics[width=\fsizefour]{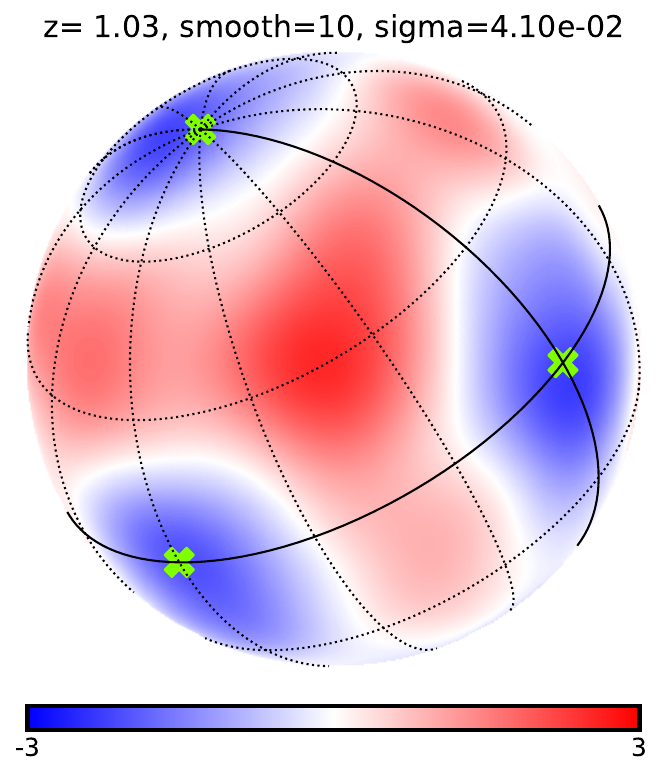}
\includegraphics[width=\fsizefour]{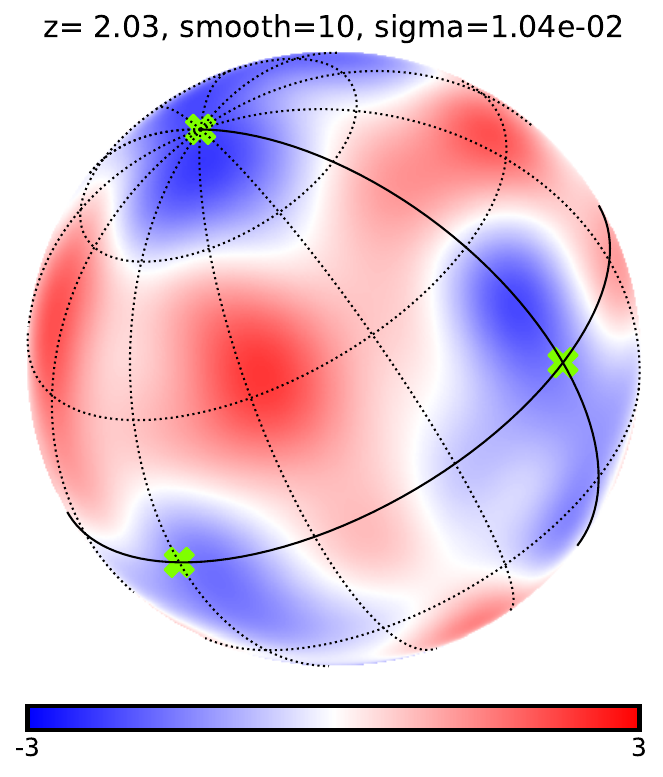}
\caption{
Similar to figure~\ref{fig:6610}, this figure presents the detected shape artifacts for the FoF halos with at least 200 particles in ELUCID simulation (first row), and for the Rockstar halos and subhalos with at least 1000 particles in Uchuu1000Pl18 simulation (second row).
}
\label{fig:elucid-uchuu}
\end{figure*}

\begin{figure*}[htbp]
\centering
\includegraphics[width=\fsizefour]{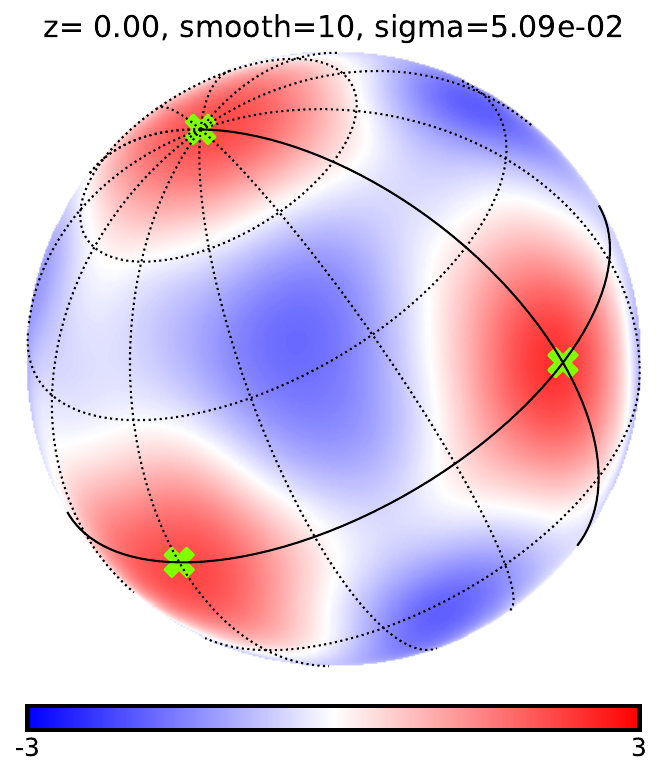}
\includegraphics[width=\fsizefour]{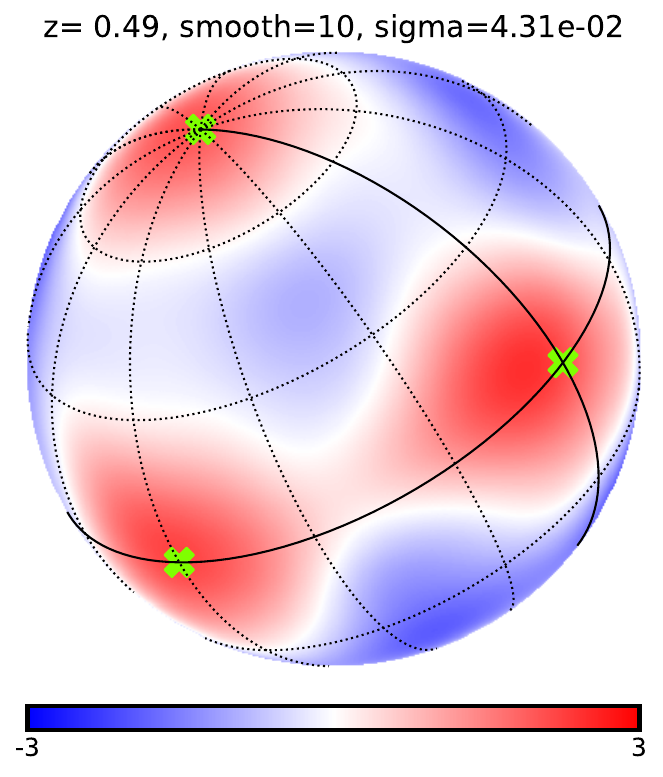}
\includegraphics[width=\fsizefour]{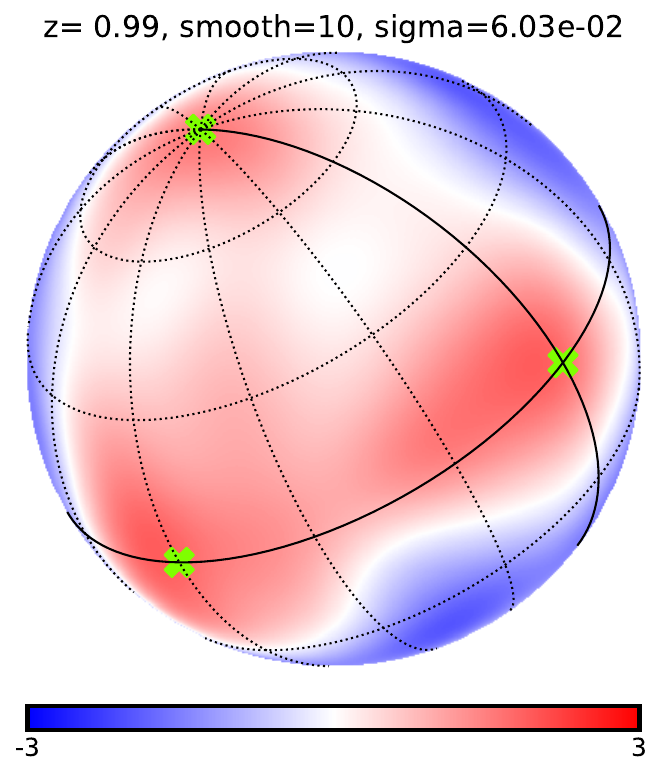}
\includegraphics[width=\fsizefour]{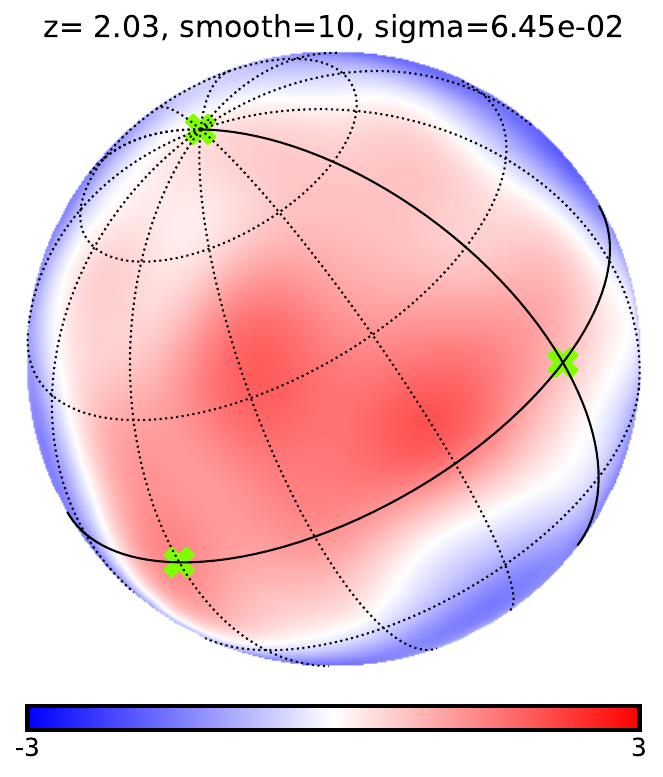}
\caption{
The shape artifacts in MDPL2 simulation.
The halo finder is \rockstar{} and the minimum halo/subhalo particle number is $100$.
Notably, the MDPL2 show opposite sign of the artifacts.
The halos and subhalos prefer to point along the three Cartesian directions for $z<1$.
}
\label{fig:mdpl}
\end{figure*}

Most of the cosmological simulation adopts grid pre-initial, as it is straightforward to implement.
Other advantages include: (i) We do not need to store the particle ID for the initial snapshot if the indices are regularly arranged.
(ii) If one want to increase or decrease the particle number by a factor of 8, $1/8$ particles have the same initial position, enabling a better control of the systematics.

In the grid pre-initial condition, particle positions are discretized as multiples of the grid size.
Consequently, proto-halo shapes are constrained to a mosaic-like structure rather than arbitrary configurations,
introducing a preferential alignment along the Cartesian axes of the simulation box.
Although subsequent gravitational evolution mitigates these artifacts through complex interactions, residual numerical effects may persist.
To robustly show these remaining numerical effects, we apply the approach described in section~\ref{sec:shape} first to the CosmicGrowth simulation.
We compute the inertia tensor for each FoF halo with $\ge 200$ member particles, and determine the major axes of halos.
By accounting the pointing of these major axes on HEALPix maps, we detect a weak preferential alignment signal as shown in figure~\ref{fig:6610}.
These HEALPix maps are normalized to zero mean and unit variance, such that the color scale directly indicates the signal-to-noise ratio.
The scatter reported in the plot title quantifies the amplitude of the residual anisotropy.

Contrary to initial expectations, the observed halo orientations do not preferentially align with the Cartesian axes of the simulation box.
Despite the low amplitude of the signal, its statistical significance is high, reaching approximately $-3\sigma$ for all three Cartesian directions across the redshift range studied ($z\le 2$).

To investigate the robustness of our finding, we analyze the ELUCID simulation, which employs the TreePM code Gadget-2 rather than the P3M method used in CosmicGrowth.
Both simulations utilize FoF halo finders.
As shown in the first row of figure~\ref{fig:elucid-uchuu}, the alignment preference in ELUCID is less pronounced than in CosmicGrowth, yet remains statistically significant for $z=0.5$ to $z=2$ (right three panels).
The underlying cause of this discrepancy is unclear, but potential explanations include
the differences in cosmological parameters (e.g., lower $\Omega_m$ and $\sigma_8$ in ELUCID), the force calculation (TreePM v.s. P3M), and subtle variations in halo-finder implementations.

To further validate our findings, we examine the Uchuu1000Pl18 simulation, which employs the GreeM code.
The force calculation is TreePM, same as ELUCID.
The halo finder is \rockstar{} and we use the halos and subhalos with $\ge 1000$ member particles.
The result is shown in the second row of figure~\ref{fig:elucid-uchuu}.
The alignment pattern remains equally distinct as observed in CosmicGrowth,
demonstrating that these numerical artifacts persist across different simulation methodologies (P3M v.s. TreePM), and alternative halo-finder algorithms (FoF v.s. Rockstar).

The final simulation we examine with grid pre-initial condtions is MDPL2.
This simulation shares the Gadget-2 code with ELUCID, and also uses the \rockstar{} halo finder as in Uchuu1000Pl18.
We modify the minimum halo size requirement to $>100$ particles to account for its relatively lower mass resolution, although it will increase the uncertainties in the halo shape measurement.
Surprisingly, we observe very different artificial pattern.
Halos exhibit a preferential alignment with the Cartesian axes for $z<1$, a reversal of the artifact pattern observed in previous simulations.
Additionally, for high redshift $z=2$, 
halo orientations display an unexpected avoidance of a certain plane.

In summary, we detect systematic alignment artifacts in halo shapes across multiple redshifts, simulation codes, and halo finders.
While the amplitude of these artifacts varies between cases, even exhibiting opposite alignment patterns in one case, their statistical signifcance remains robust.
When applying higher mass thresholds, the reduced sample size diminishes the clarity of the pattern (not shown here), though their underlying significance persists.

\subsection{Tracing the Shapes across Time}
\label{sec:evolution}

\begin{figure*}[htbp]
\centering
\includegraphics[width=\fsizefour]{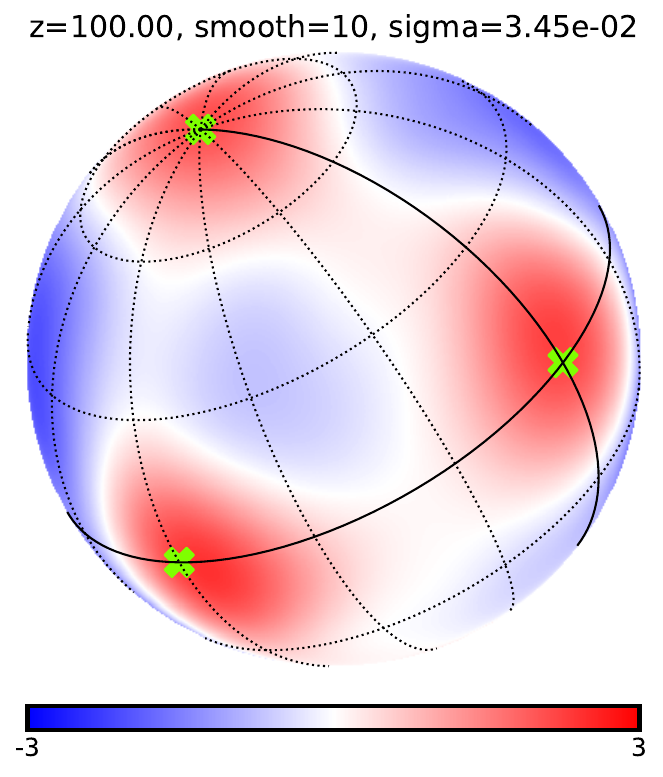}
\includegraphics[width=\fsizefour]{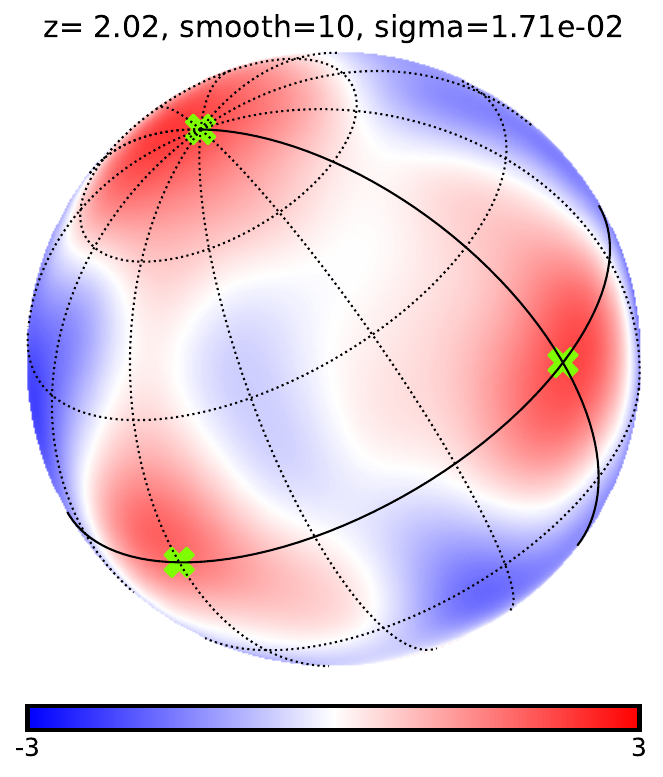}
\includegraphics[width=\fsizefour]{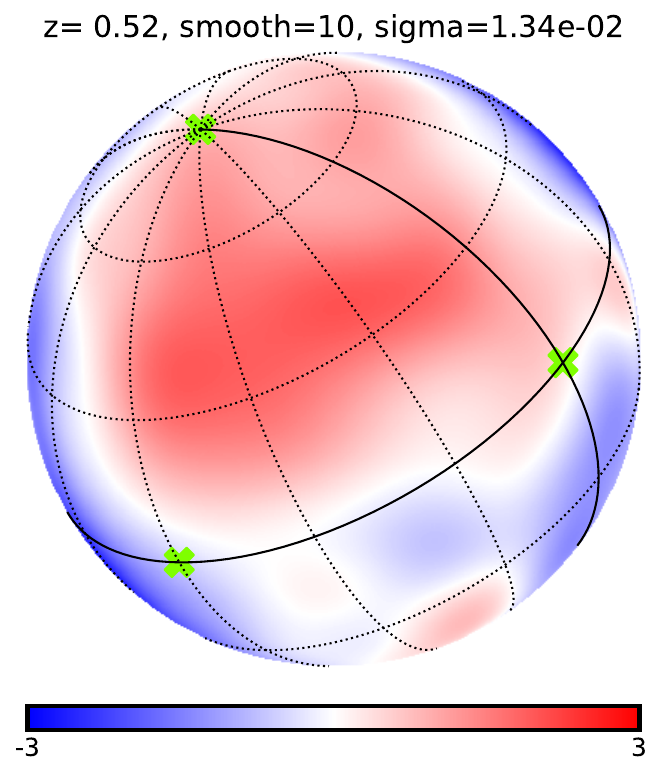}
\includegraphics[width=\fsizefour]{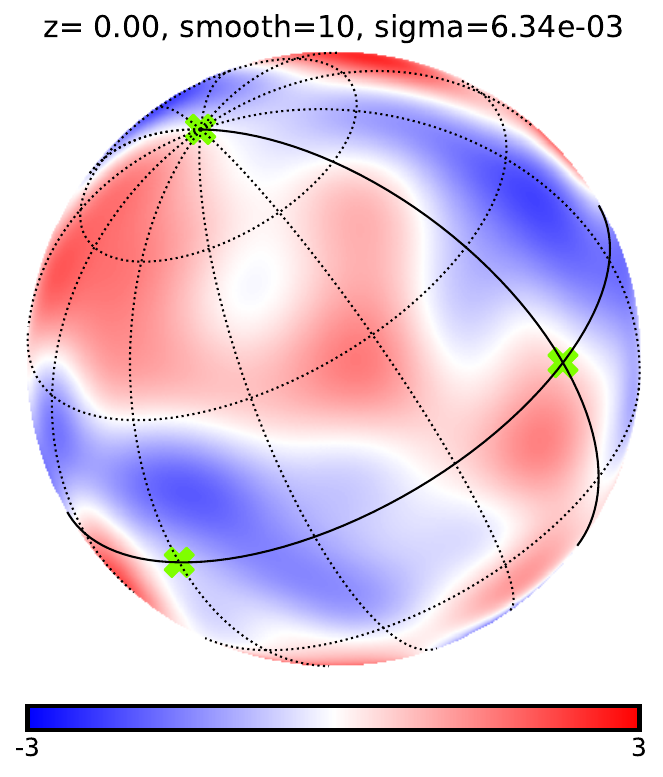}
\\
\includegraphics[width=\fsizefour]{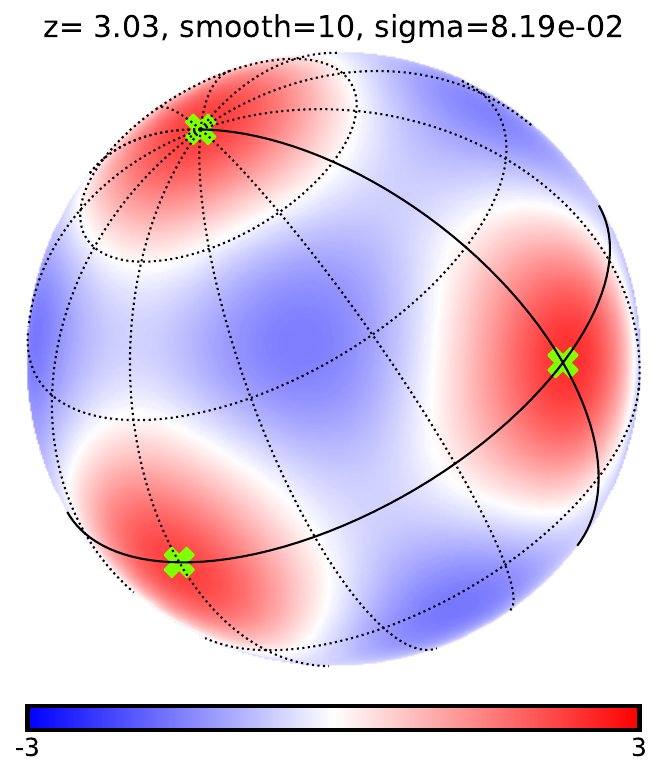}
\includegraphics[width=\fsizefour]{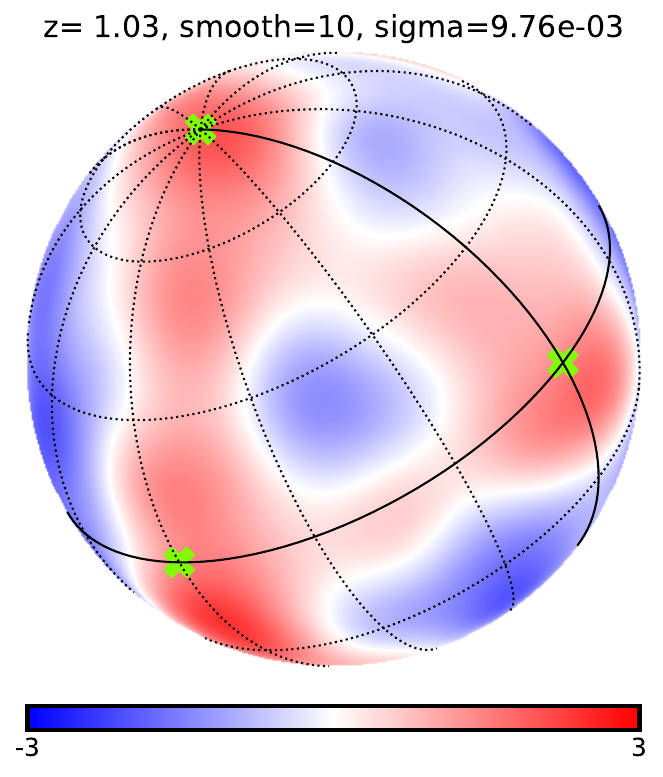}
\includegraphics[width=\fsizefour]{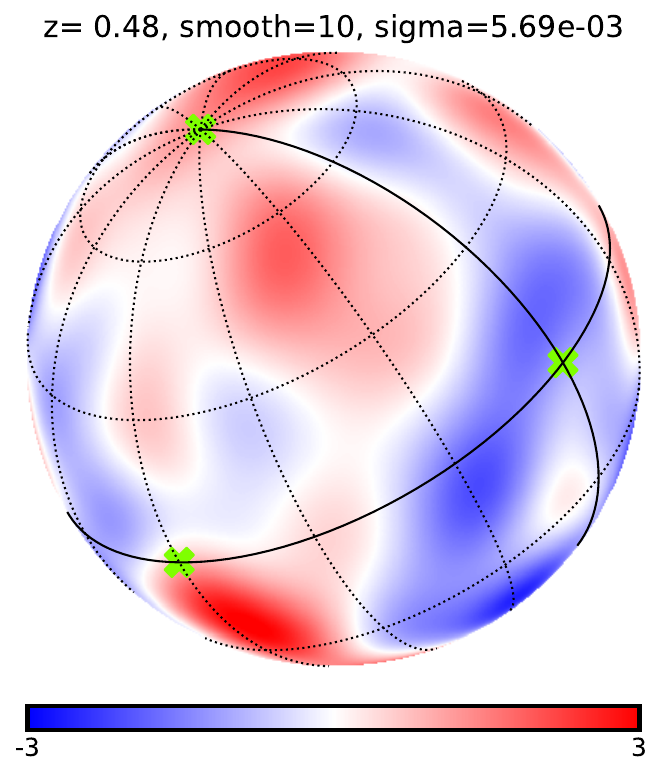}
\includegraphics[width=\fsizefour]{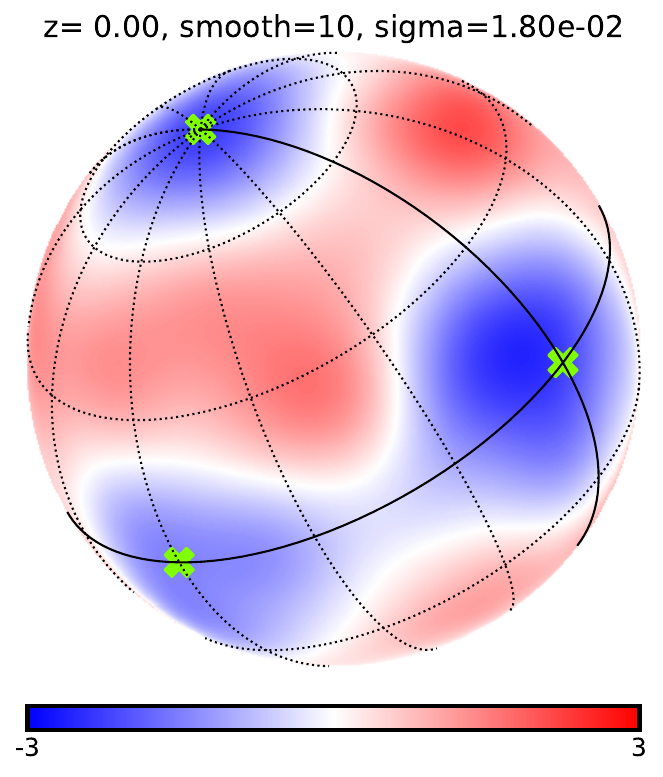}
\caption{
\emph{First row:} Evolution of the halo particles orientation preference in the ELUCID simulation, traced from initial condition at $z=100$ to $z=0$ (left to right).
For $z\ge 2$, these halo particles show preferential alignment with Cartesian axes,
while below $z\approx 0.5$, this preference disappears.
\emph{Second row:} The CosmicGrowth simulation exhibits a clear flipping behavior in the alignment preference, with $z\ge 1$ the alignment preferring to point at Cartesian axes, and for $z=0$ the alignment avoiding to point at Cartesian axes.
The transition occurs around $z=0.5$.
In figure~\ref{fig:elucid_minor} the same flipping behavior is observed for minor axes in ELUCID.
}
\label{fig:elucid-6610-evolve}
\end{figure*}

Given the unexpected behavior of these artifacts, a straightforward idea is to directly investigate the artifacts in proto halos, i.e. the initial distribution of particles belonging to late-time halos.
In addition, we can trace the particle positions at multiple redshifts to see the evolution of the artifacts.

Using ELUCID, which shows almost no detection of the artifacts in halos today,
the proto halos indeed have the alignment pattern with expectation.
As shown in the first panel in figure~\ref{fig:elucid-6610-evolve},
they prefer to align with the Cartesian axes,
with an amplitude of $\sim 2.5\sigma=0.083$.
This preference clearly persists until $z=2$, as shown in the second panel.
Below $z=0.5$, the preference disappears.

To cross validate this behavior, we perform the same analysis using CosmicGrowth.
In the second row of figure~\ref{fig:elucid-6610-evolve}, the alignment pattern is shown for redshift $z=3$ to $z=0$ (left to right).
For $z\ge 1$, the halo particles prefer to align with Cartesian axes of the simulation box, while for $z=0$, the preference reverses.
The flipping occurs between $z=1$ and $z=0$, leading to no detection at $z=0.5$.
In appendix~\ref{sec:appendix_minor}, we find that the same flipping behavior occurs for the minor axes in ELUCID, between $z=2$ and $z=0$.

From the above analysis, we confirm that the artifacts in halo shapes stem from the initial condition, and slowly decrease in amplitude in the sequential evolution, flip sign at intermediate redshift ($z\sim 0.5$), and become prominent again today.

\subsection{Filament Orientation}
\label{sec:filaorien}

\begin{figure*}[htbp]
\centering
\includegraphics[width=\fsizefour]{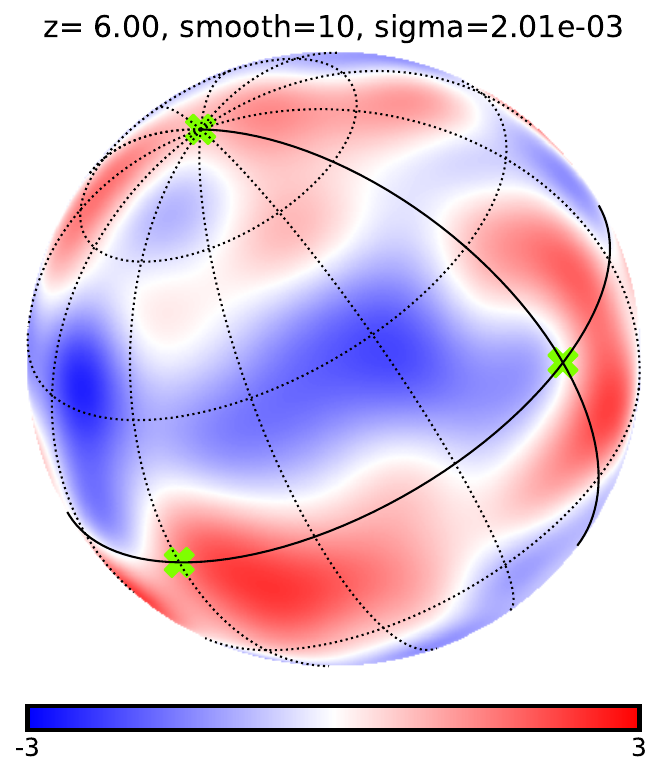}
\includegraphics[width=\fsizefour]{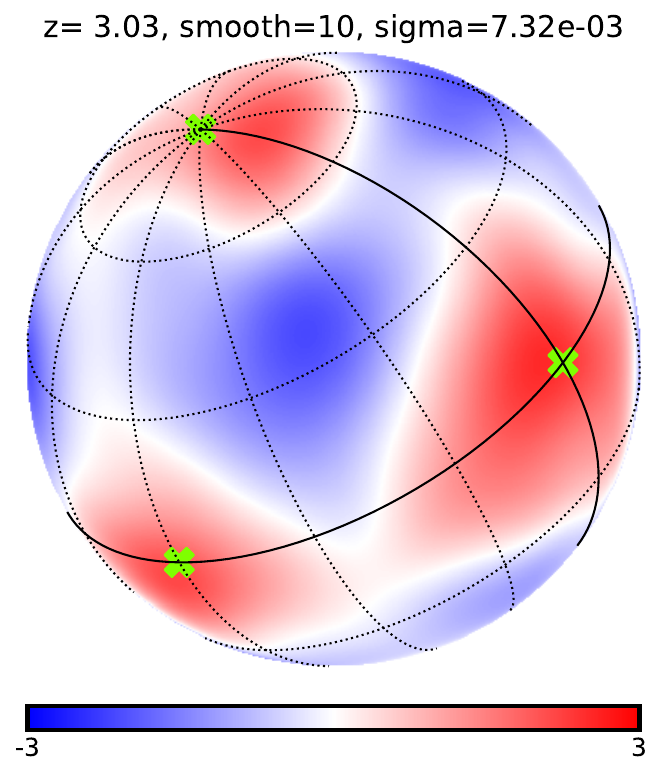}
\includegraphics[width=\fsizefour]{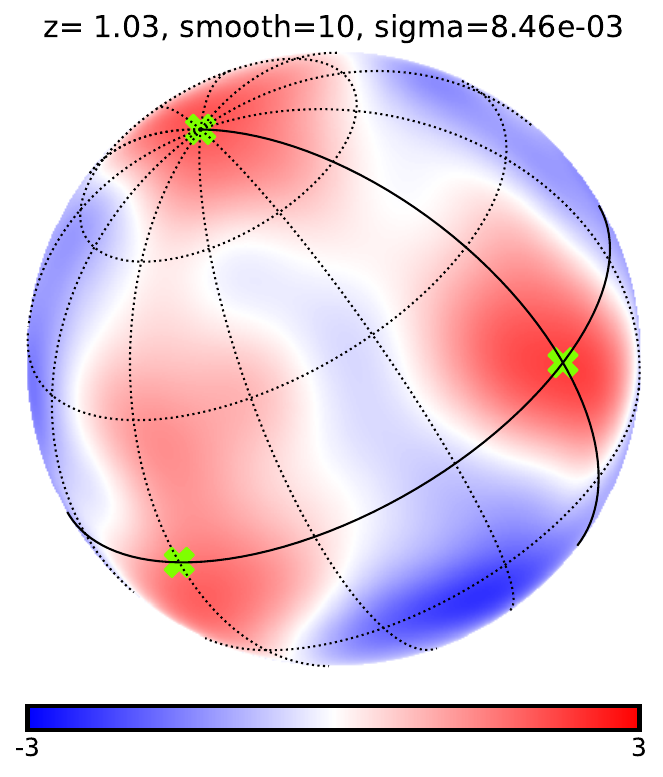}
\includegraphics[width=\fsizefour]{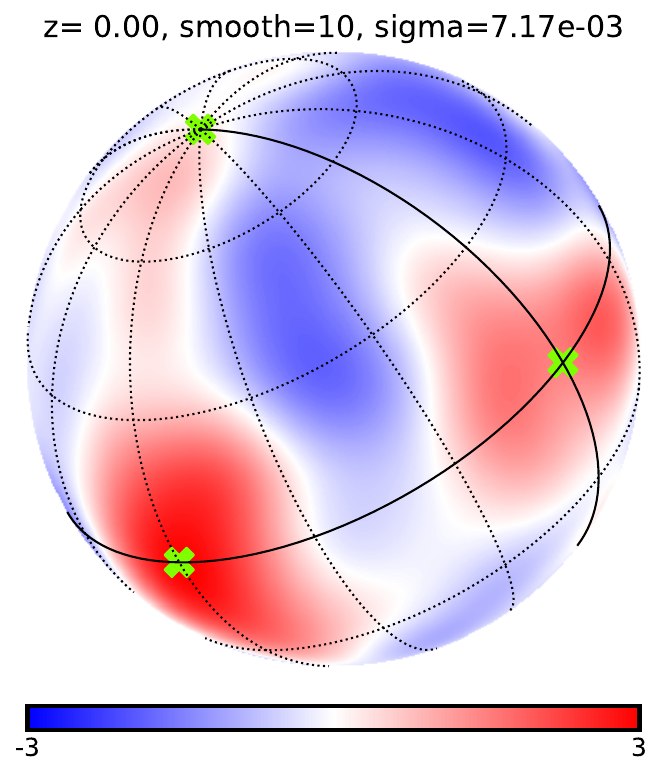}
\caption{
In CosmicGrowth simulation, the filaments also show clear orientation preference along the Cartesian axes for $z\le 6$.
This behavior is most prominent in $z=3$ and $z=1$ plot. 
}
\label{fig:6610-fila}
\end{figure*}

Since the halo formation and merging is strongly related to the large-scale structure,
we guess the flipping behavior is related to the filaments and the halo behavior inside the filaments. 
So we adopt the approach describe in section~\ref{sec:fila} to measure the preference of filament orientation.

Obviously, in figure~\ref{fig:6610-fila} filaments also show orientation preference, pointing along the Cartesian axes, same as the proto halos.
Because of the anisotropy in the initial distribution, matter in intermediate and minor axis directions will collapse earlier, leaving the filament being 1D overdensity structures and the pointing has the same preference.
Different from halos, the preference is always in the same direction across cosmic time.
The amplitude seems to increase from $z=6$ to $z=3$, and decrease from $z=1$ to $z=0$.
The low significance at high redshift is partially due to the limited number of identified filaments.

Therefore, at least phenomenologically, the halo particles initially prefer to align with filaments, and then align out of filaments at low redshift.
The alignment preference in the early phase is probably due to the fact that both halo particles and filaments are evolving under gravity together.
The filaments orientation persists well once they are formed, while the halos will experience further accretion and merging.


\subsection{Glass Pre-initial Loading}
\label{sec:kun}

\begin{figure*}[htbp]
\centering
\includegraphics[width=\fsizefour]{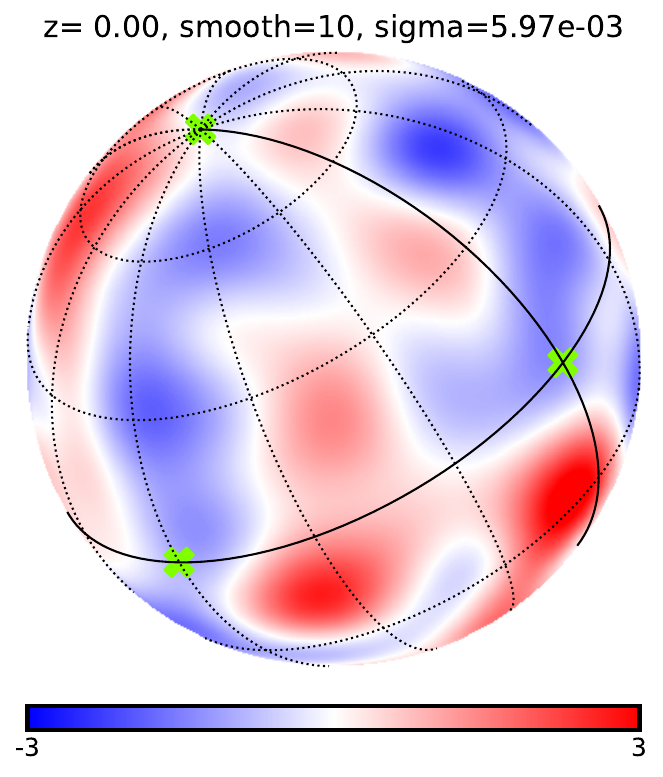}
\includegraphics[width=\fsizefour]{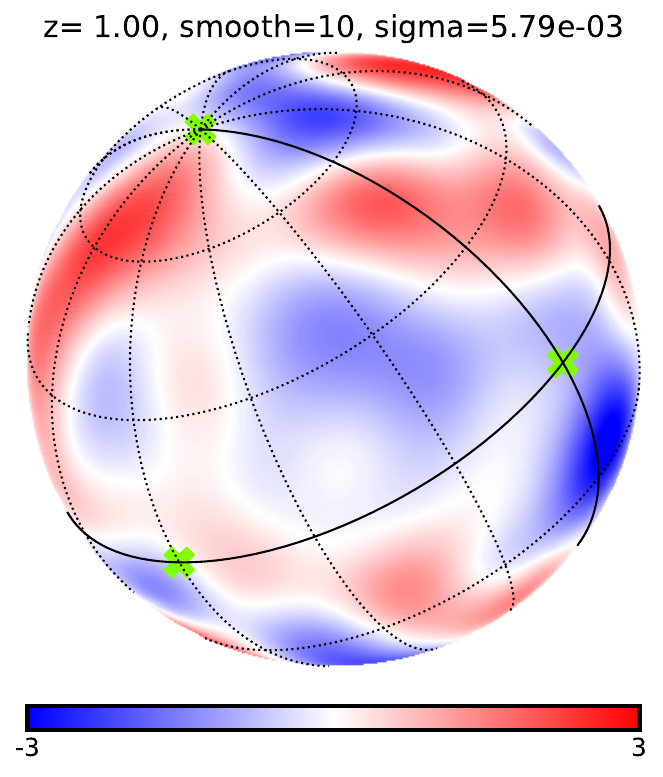}
\includegraphics[width=\fsizefour]{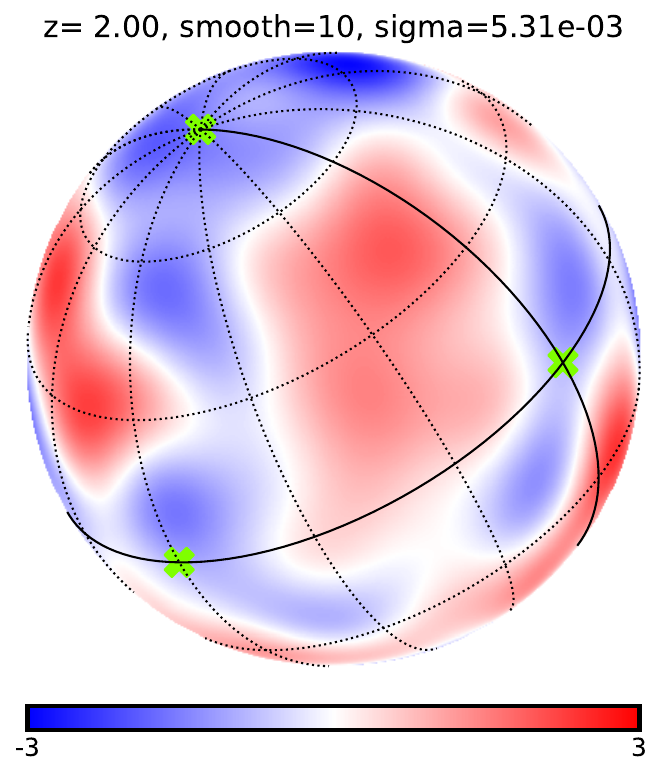}
\includegraphics[width=\fsizefour]{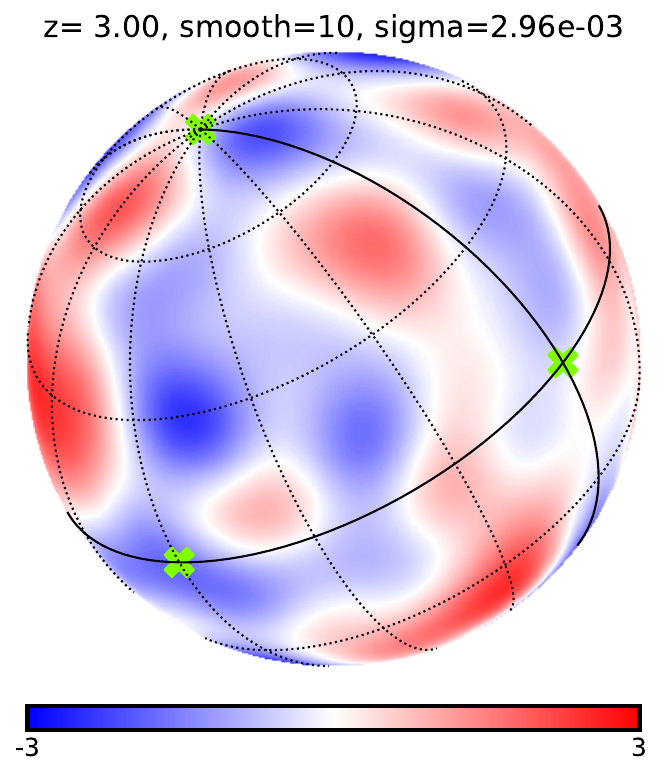}
\\
\includegraphics[width=\fsizefour]{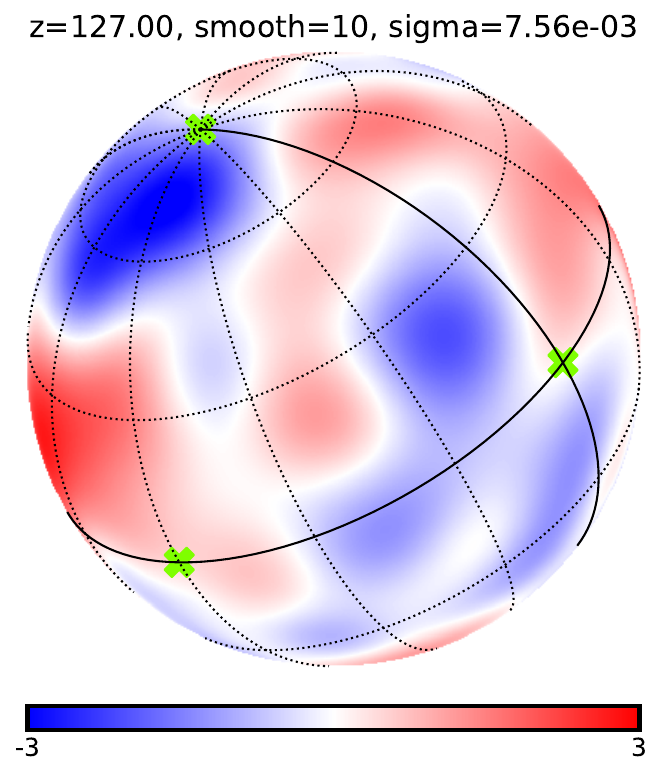}
\includegraphics[width=\fsizefour]{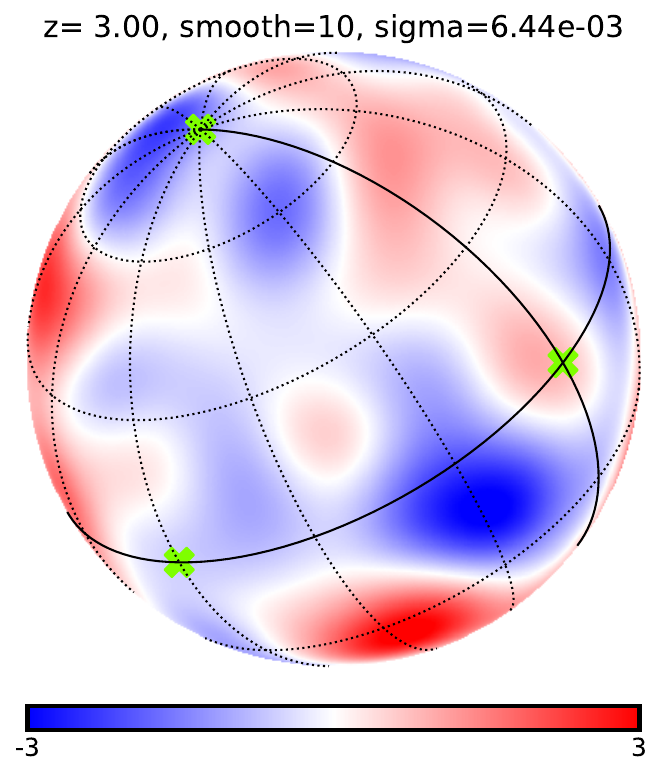}
\includegraphics[width=\fsizefour]{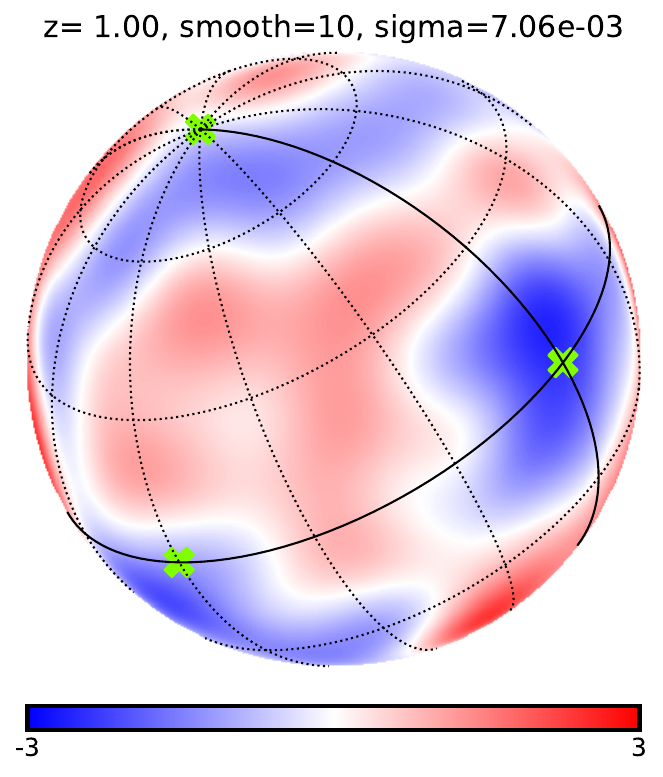}
\includegraphics[width=\fsizefour]{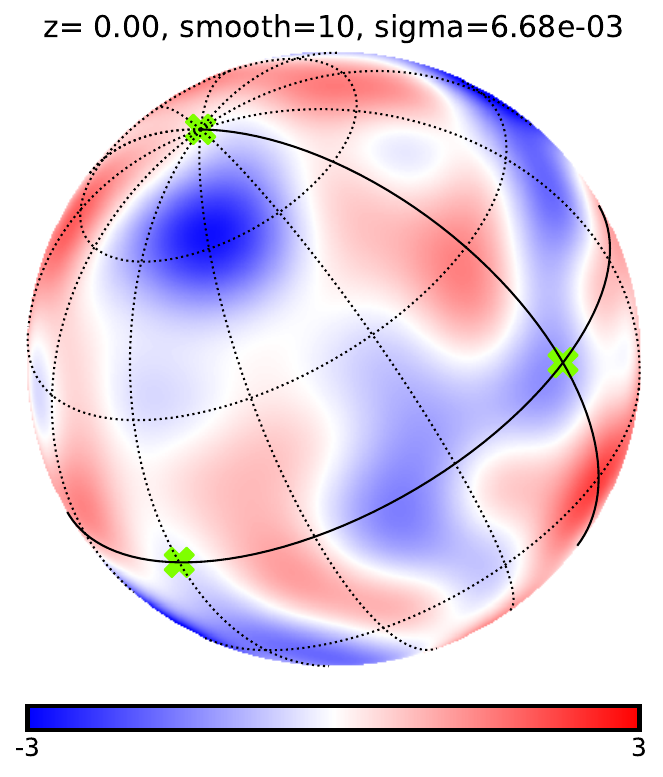}
\\
\includegraphics[width=\fsizefour]{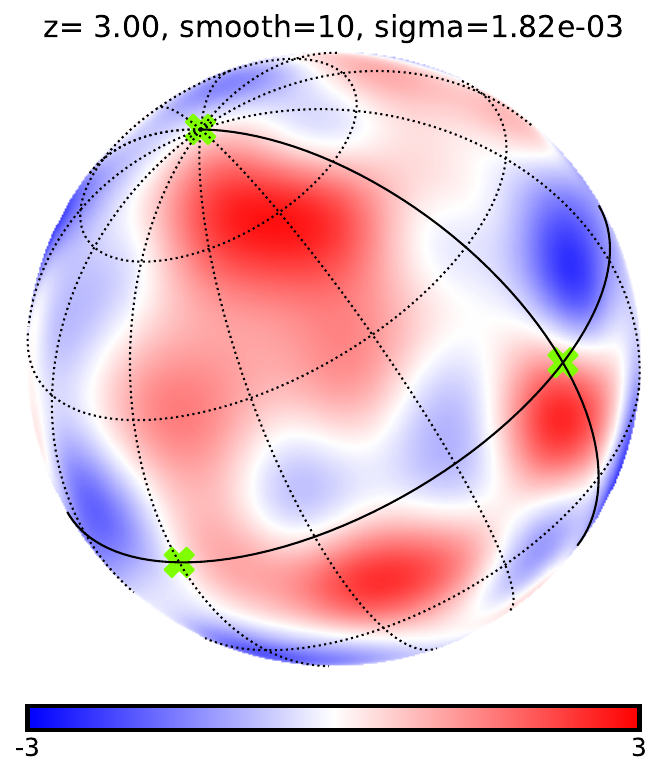}
\includegraphics[width=\fsizefour]{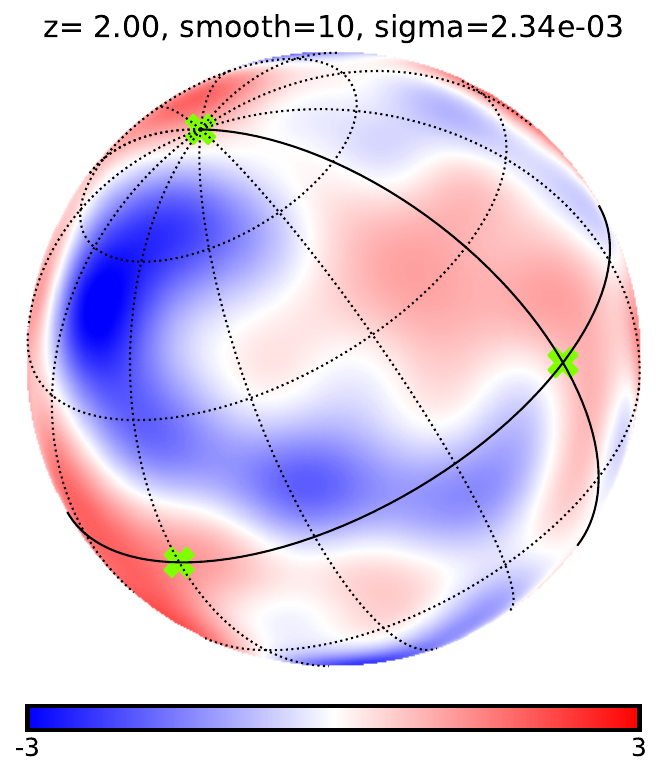}
\includegraphics[width=\fsizefour]{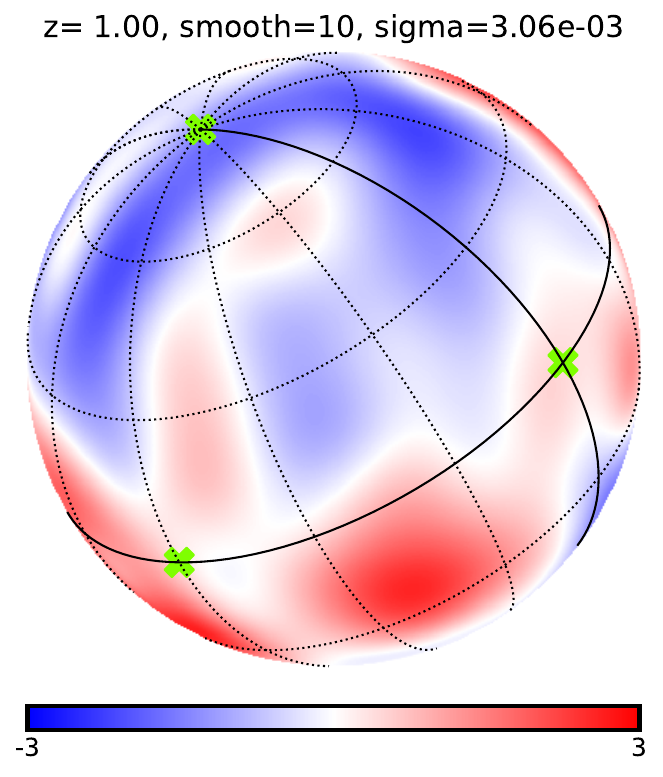}
\includegraphics[width=\fsizefour]{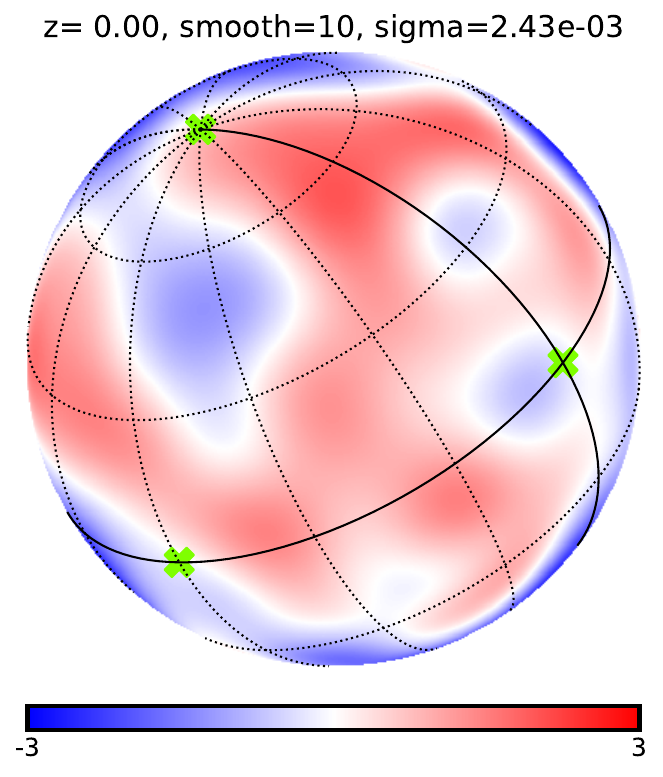}
\caption{
\emph{First row:} No shape artifacts are observed in Kun simulation for halos in multiple redshifts.
\emph{Second row:} Backward tracing the halo particles, no shape artifacts are observed in initial conditions (left panel), either in the subsequent evolution (right three panels).
\emph{Third row:} No filament orientation preference is observed in Kun simulation.
}
\label{fig:kun-all}
\end{figure*}

To determine whether the alignment preferences observed earlier are artifacts of grid-type initial conditions, we conduct the same analysis in section~\ref{sec:z-dependence} to \ref{sec:filaorien} on Kun fiducial simulation.
Different from the above simulations, Kun adopts glass-type pre-initial condition.
Thus, we expect that Kun fiducial simulation is inherently free of grid-induced artifacts.
From the results in figure~\ref{fig:kun-all},
no alignment preference is detected in halo shapes (first row), historical positions of halo particles (second row), and filament orientations (third row).
This result confirms that the alignment patterns in CosmicGrowth, ELUCID, Uchuu1000Pl18, and MDPL2 arise solely from grid-type initial conditions. 
Moreover, the sequential evolution process does not introduce such numerical artifacts.

\subsection{Excess Probability of Alignments}
\label{sec:cosineexcess}

\begin{figure*}[htbp]
\centering
\includegraphics[width=7cm]{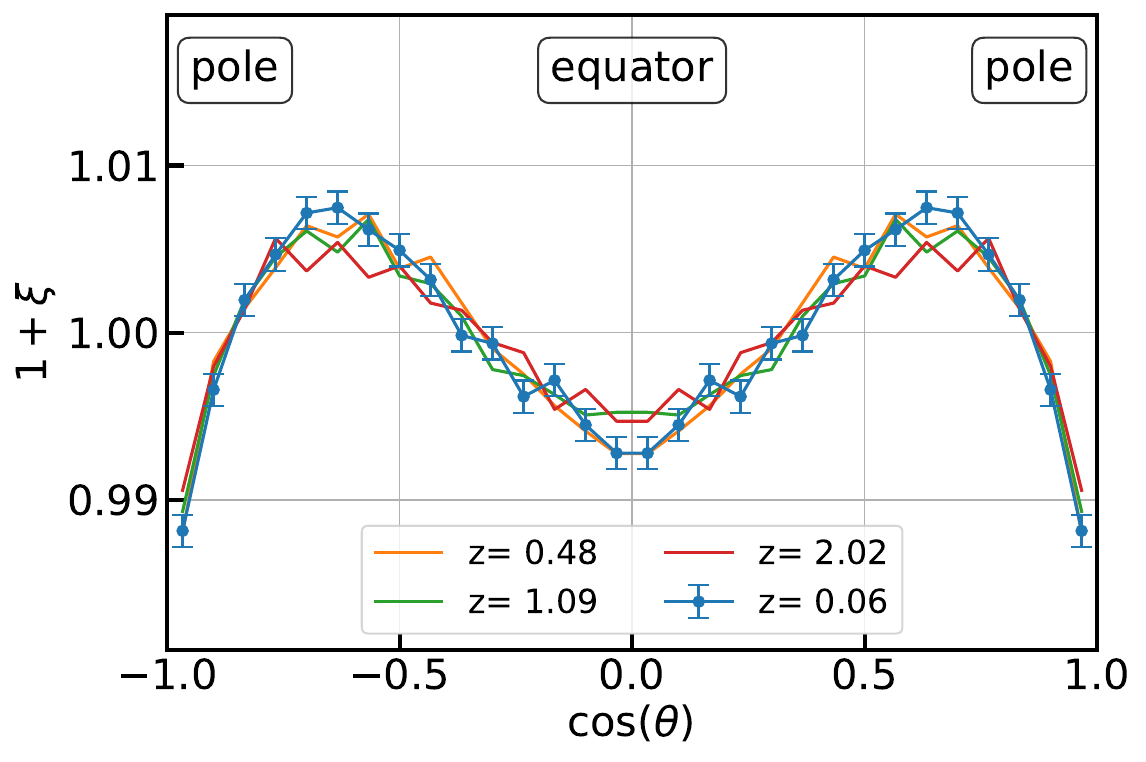}
\includegraphics[width=7cm]{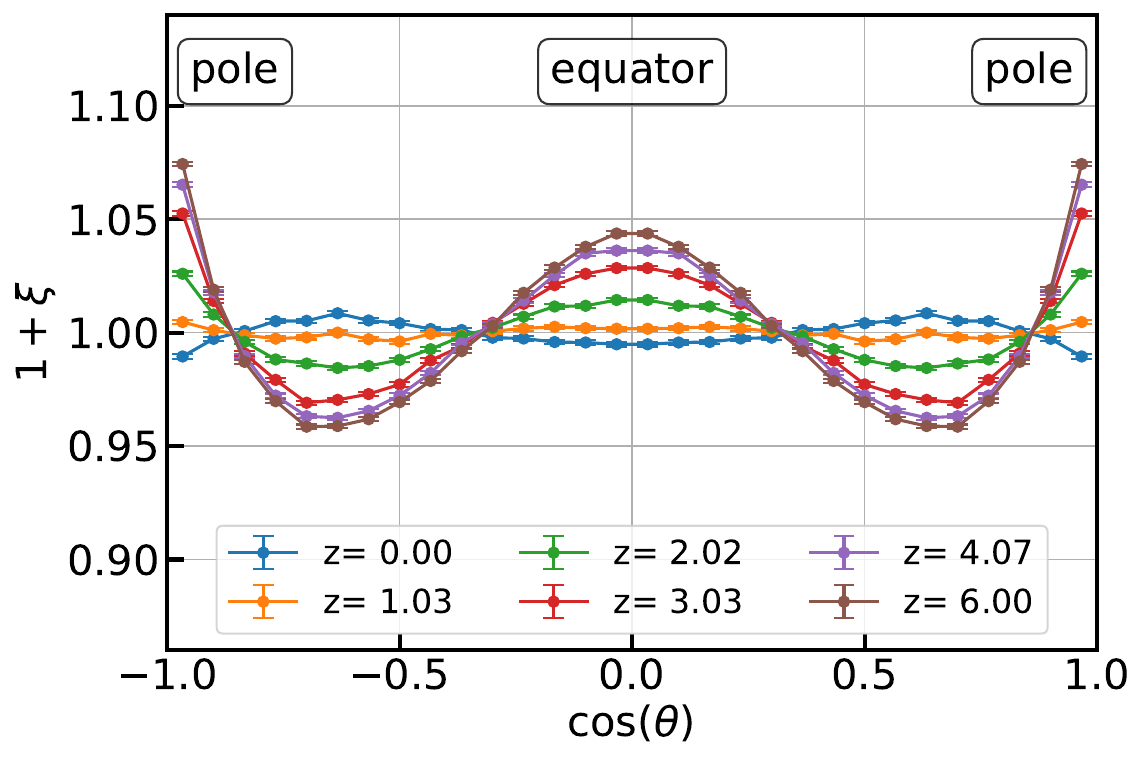}
\caption{
\emph{Left panel:} The excess probability of the cosine angle between the halo major axes and the three Cartesian axes of the simulation box.
\emph{Right panel:} The time evolution of the alignment signal by tracing the past position of particles in $z=0$ halos.}
\label{fig:cosineexcess}
\end{figure*}

While the visual inspection in previous sections provides a straightforward qualitative assessment, we now quantify the alignment signal by computing the excess probability described in section~\ref{sec:visual}.
The result for CosmicGrowth simulation is shown in figure~\ref{fig:cosineexcess}.
Taking $z$-axis as the reference direction,
$\cos(\theta)=\pm 1$ indicates alignment with the $z$-axis, 
and $\cos(\theta)=0$ represents alignment with the directions in the $x$-$y$ panel, including the $x$-axis and $y$-axis.
Thus, the alignment signal has the same sign in $\cos(\theta)=-1, 0, 1$, but the amplitude for $\cos(\theta)=0$ is low.
Note that we stack the excess probability calculated relative to three Cartesian axes together.
Specifically, we measure alignment signals of $-1\%, 0.8\%$ and $-0.5\%$ for $|\cos(\theta)|=1, 0.7$ and $0$, respectively.
These amplitudes remain stable across redshifts ($z\sim 2$ to $z\sim 0$), with Poisson error bars overplotted for the $z=0.06$ results.
Despite the modest amplitude ($\sim 1\%$), the signal is statistically significant and consistent with the visual trends in section~\ref{sec:z-dependence}.

The $1\%$ level alignment preference seems not to be fully negligible compared to the excess probability measurements in \cite{Codis15,Codis18}, in which a few percent level of alignment is observed between galaxy spins and filaments.
Note that their study probed alignments between two astrophysical directions, whereas ours measures alignment relative to the simulation box’s Cartesian axes. 
Consequently, if both vectors in a pairwise analysis exhibit a $1\%$ preference for the Cartesian axes, their mutual artificial alignment would be suppressed to $0.01\%$, effectively negligible.

Figure~\ref{fig:cosineexcess} also traces the evolution of halo particle alignment from $z=6$ to $z=0$.
The signal displays a monotonic trend.
The amplitude is strong at early times, weakening with decreasing redshift, flipping sign near $z\sim 1$, and rebounding at $z\sim 0$.
Crucially, the zero-crossing point remains consistent throughout, suggesting that the low- and high-redshift patterns share a common origin, rather than arising from distinct numerical artifacts introduced at early and late times.
The result for minor axes is shown in appendix~\ref{sec:appendix_minor}, in which a strong amplitude is observed at early times, as the minor axes are by definition shorter than major axes, and thus more sensitive to the discretizations.

\subsection{Possible Reason for Low Redshift Artificial Pattern}
\label{sec:flipreason}

In figure~\ref{fig:6610}, the CosmicGrowth halos at $z=0.5$ and $z=1$ show alignment preference away from the Cartesian axes.
In contrast, the second row of figure~\ref{fig:elucid-6610-evolve} shows that the particles in $z=0$ halos have no alignment preference at $z=0.5$, and show preference towards the Cartesian axes at $z=1$.
This implies that the well-formed halos have distinct alignment behavior from the particles which are forming halos.
Therefore, the low redshift alignment pattern is related to the formation of halos.

Halos form through accretion and mergers, so their residual numerical artifacts likely reflect the cumulative artifacts of their progenitors.
If these artifacts combine statistically,
an excess alignment probability along $(1,0,0)$ and $(0,1,0)$ could produce a resultant excess along $(1,1,0)$.
To illustrate, consider Gaussian-distributed variables 
$X\sim\mathcal{N}(2.5,2)$ and $Y\sim\mathcal{N}(7.5,2)$.
Their average $Z=\frac12(X+Y)$ follows $\mathcal{N}(5,\sqrt{2})$,
shifting the orignal probability excess regions.

However, this model is overly simplistic.
Halos typically form via multiple mergers, so we hypothesize that 
unmerged halos retain shape artifacts aligned with the Cartesian axes, while merged halos exhibit more complex distribution.
The no detection of alignment preference for halo particles at $z=0.5$ (the third plot in the second row of figure~\ref{fig:elucid-6610-evolve}) is actually a combination of multiple components: (i) the well-formed halos after merger (alignment preference away from Cartesian axes),
(ii) the recently formed halos without merger history (alignment with Cartesian axes),
(iii) the particles will be accreted on halos during $z=0.5$ and $z=0$,
and (iv) the particles will form new halos between $z=0.5$ and $z=0$ (align preference with Cartesian axes).
A rigorous test would require constructing merger trees and categorizing halos by merger frequency.
We defer this analysis to future work.

\section{Conclusions and discussion}
\label{sec:conclusion}

In this work, we systematically investigated numerical artifacts in halo shapes across multiple cosmological simulations, including CosmicGrowth, ELUCID, Uchuu1000Pl18, MDPL2 and Kun fiducial.
These simulations span diverse configurations, including  both P3M and TreePM force calculation methods, various simulation codes, both FoF and Rockstar halo finder,
and grid-type or glass-type pre-initial conditions.
Our key findings reveal a statistically significant alignment pattern in simulations with grid-type initial conditions, albeit with low absolute amplitude.
The halo shapes exhibit a preference for not aligning with Cartesian axes of the simulation box.

By tracing the particle positions back in time, we observed that the proto-halo shapes do align with Cartesian axes simply due to the discrete effect in grid-type initial condition.
However, the alignment pattern reverses at low redshift.
In contrast, filaments always show a stable alignment signal with Cartesian axes.

Despite the persistence of these artifacts, their amplitude ($\sim 1\%$) is unlikely to significantly bias intrinsic alignment measurements in mock catalogs (e.g., \cite{Zhang25}), even assuming perfect alignment between galaxy and halo shapes.
However, our work provides another example that glass-type pre-initial condition is optimal than grid-type, by investigating the anisotropic statistics like ref.\cite{Masaki21}.

One straightforward future work is to understand the mechanism of alignment pattern reversal. The redshift-dependent sign flip in halo alignment has not yet been fully explained and warrants further study.
The MDPL2 simulation exhibits alignment patterns similar to other grid-type simulations but with inverted signs.
We hypothesize this also stem from its pre-initial conditions, and plan to check its initial condition generator and dark matter particle trajectories.

\appendix
\section{Artifacts in Minor Axes}\label{sec:appendix_minor}

\begin{figure*}[htbp]
\centering
\includegraphics[width=\fsizefour]{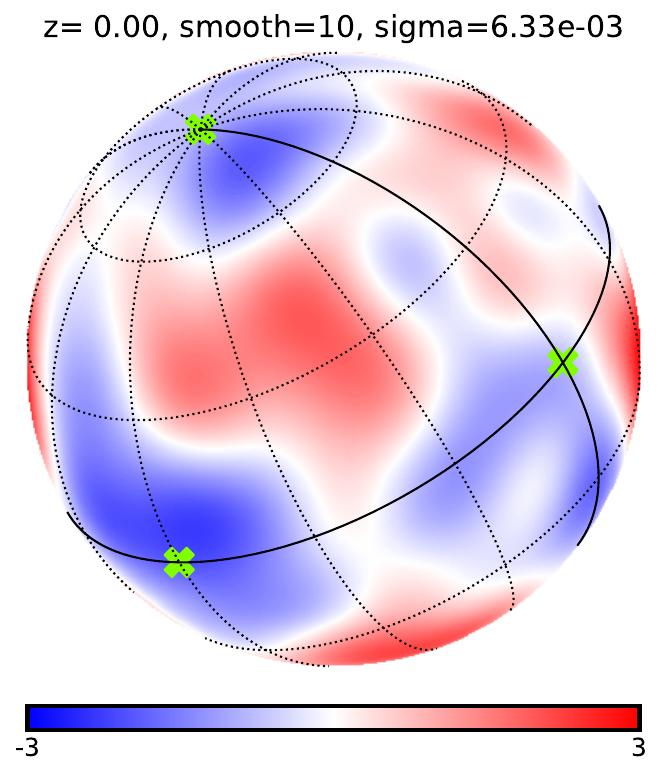}
\includegraphics[width=\fsizefour]{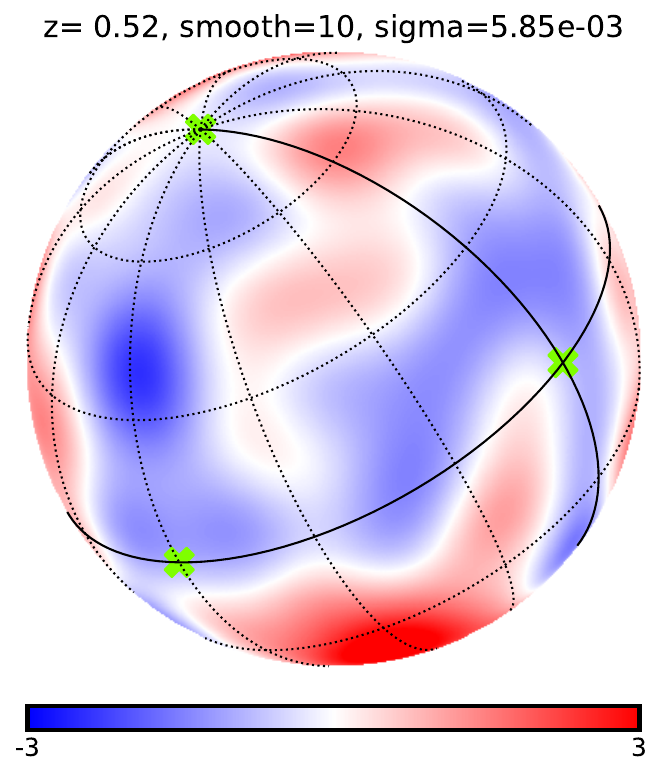}
\includegraphics[width=\fsizefour]{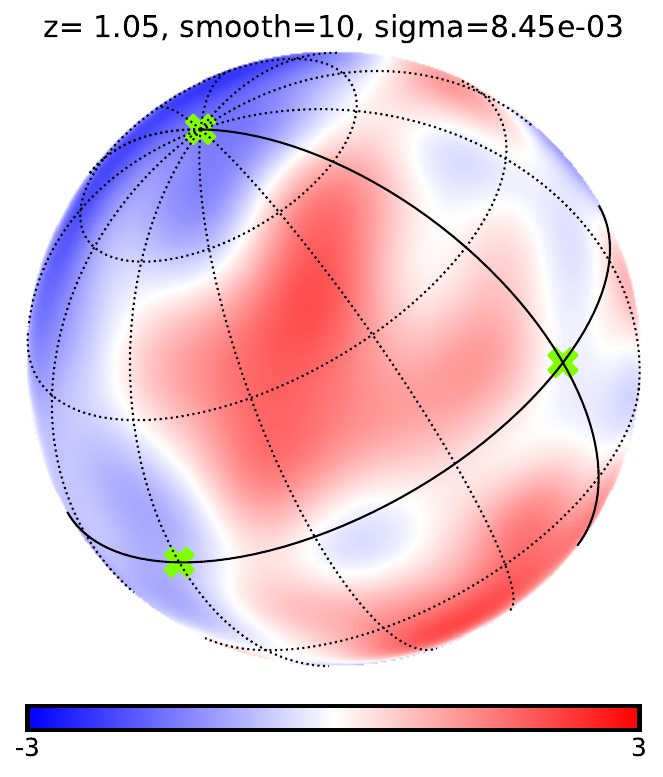}
\includegraphics[width=\fsizefour]{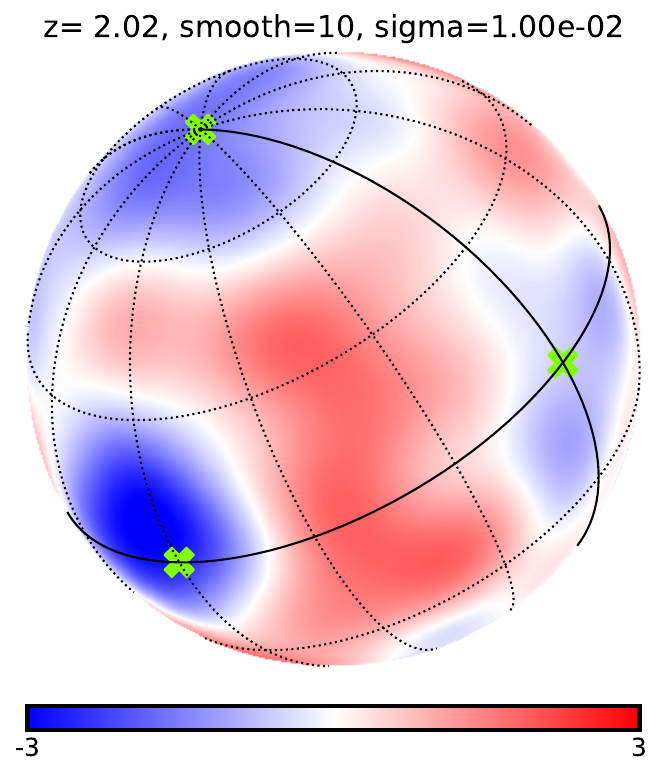}
\\
\includegraphics[width=\fsizefour]{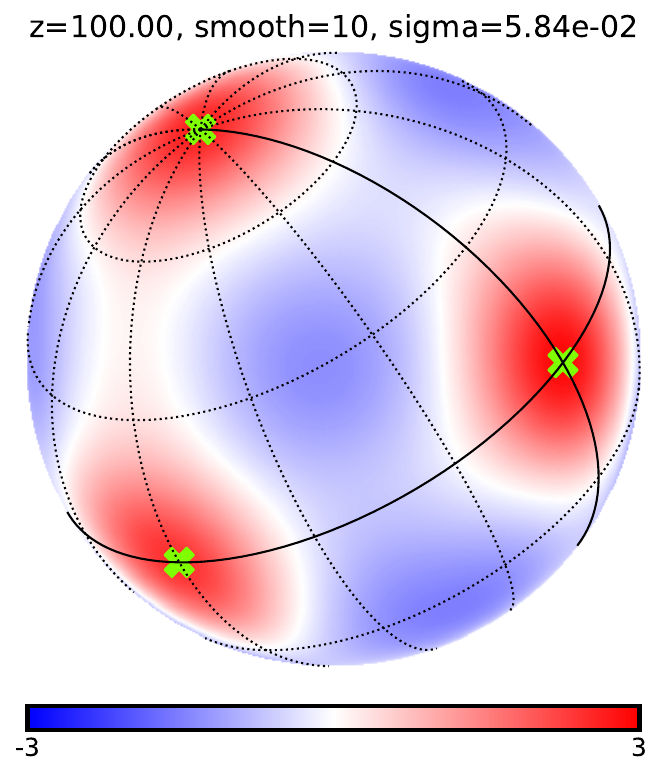}
\includegraphics[width=\fsizefour]{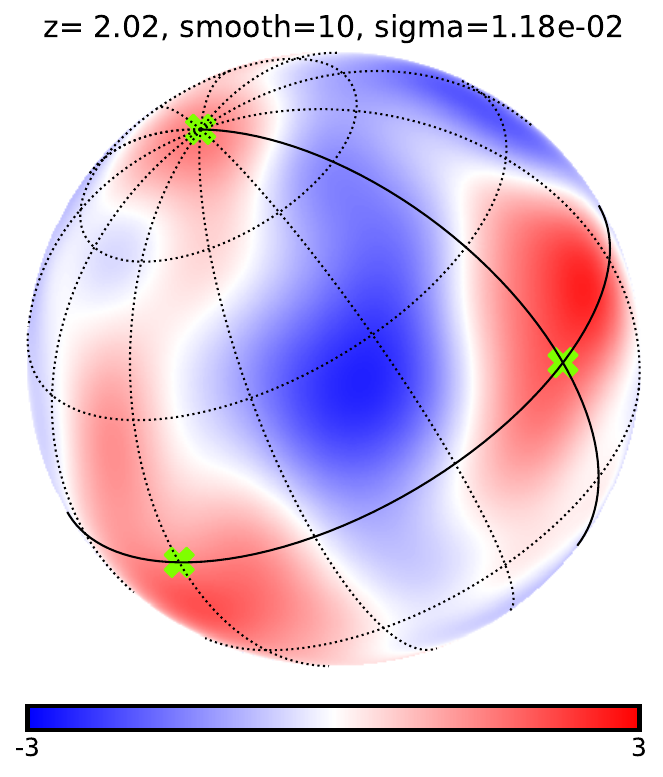}
\includegraphics[width=\fsizefour]{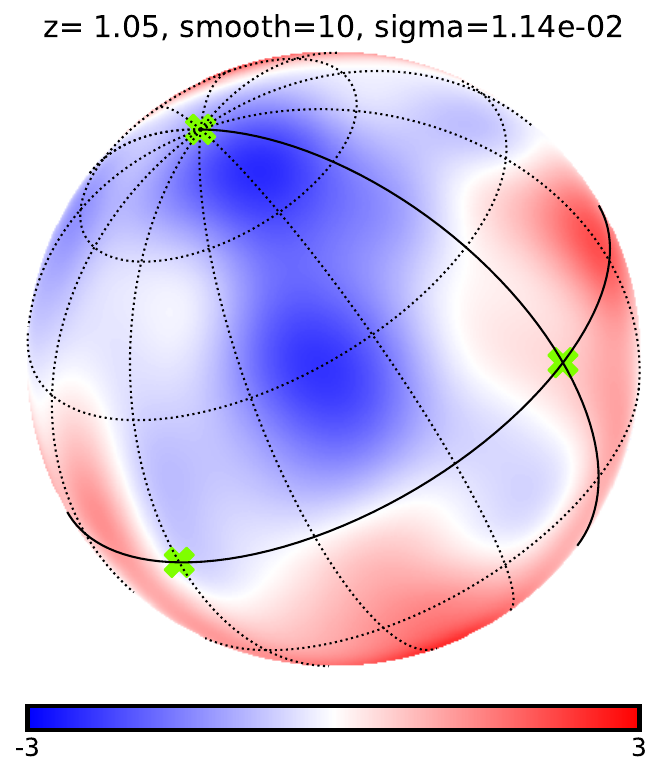}
\includegraphics[width=\fsizefour]{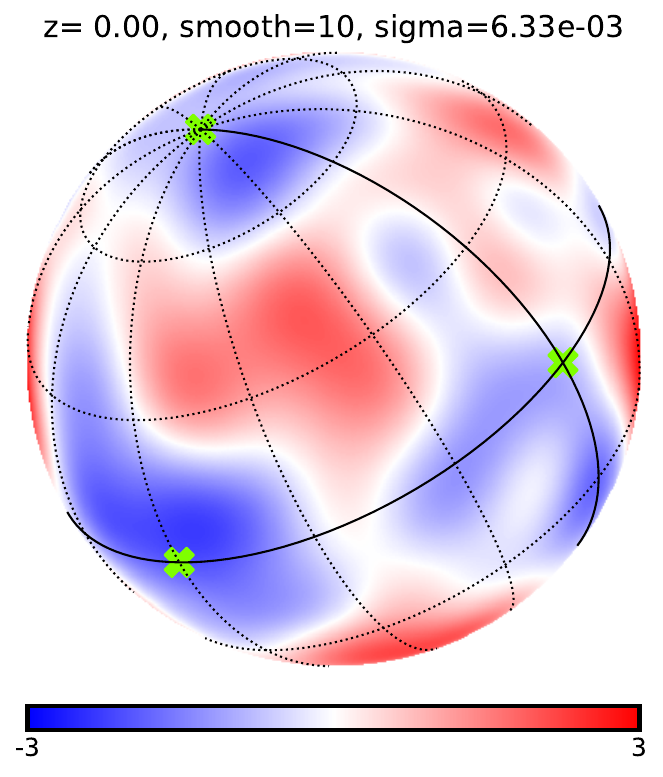}
\caption{
Artifacts from minor axes in ELUCID simulation.
\emph{First row:} Minor axes show numerical artifacts at $z=0$ (first panel), while major axes have no detection as shown in the first panel of figure~\ref{fig:elucid-uchuu}.
Both major and minor axes show artifacts at $z=2$.
\emph{Second row:} A clear alignment preference flipping is observed between $z=2$ and $z=0$ for minor axes.
}
\label{fig:elucid_minor}
\end{figure*}

\begin{figure*}[htbp]
\centering
\includegraphics[width=7cm]{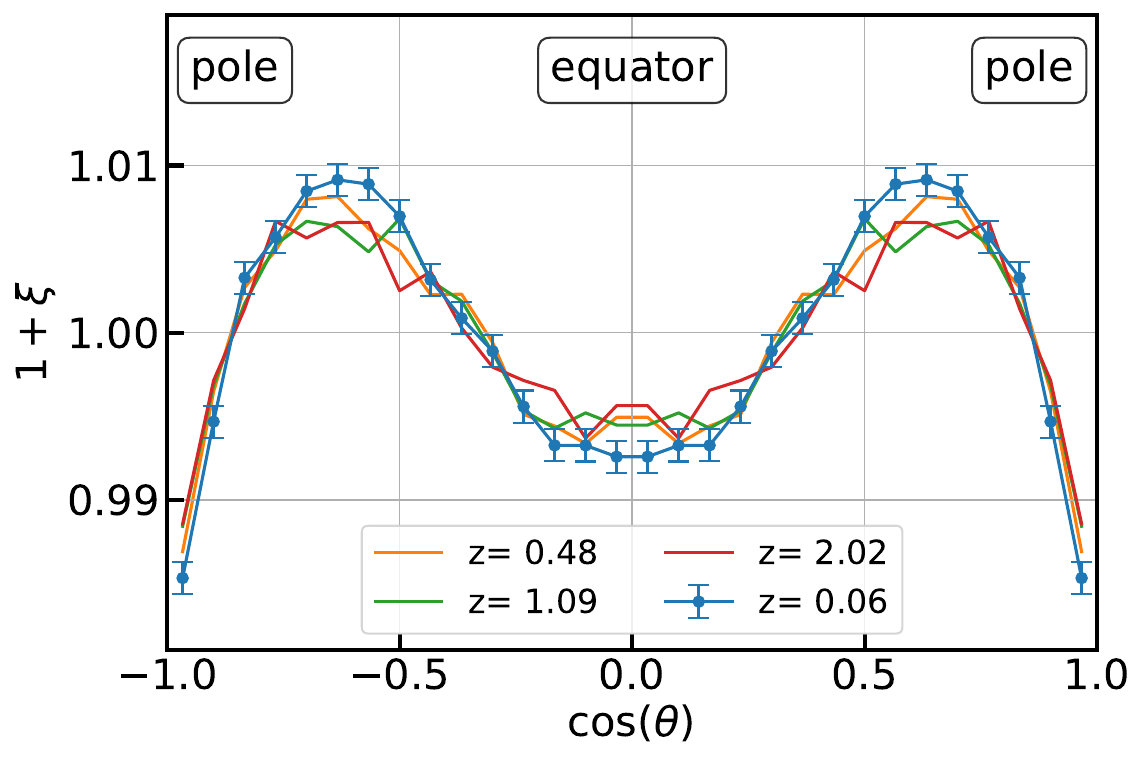}
\includegraphics[width=7cm]{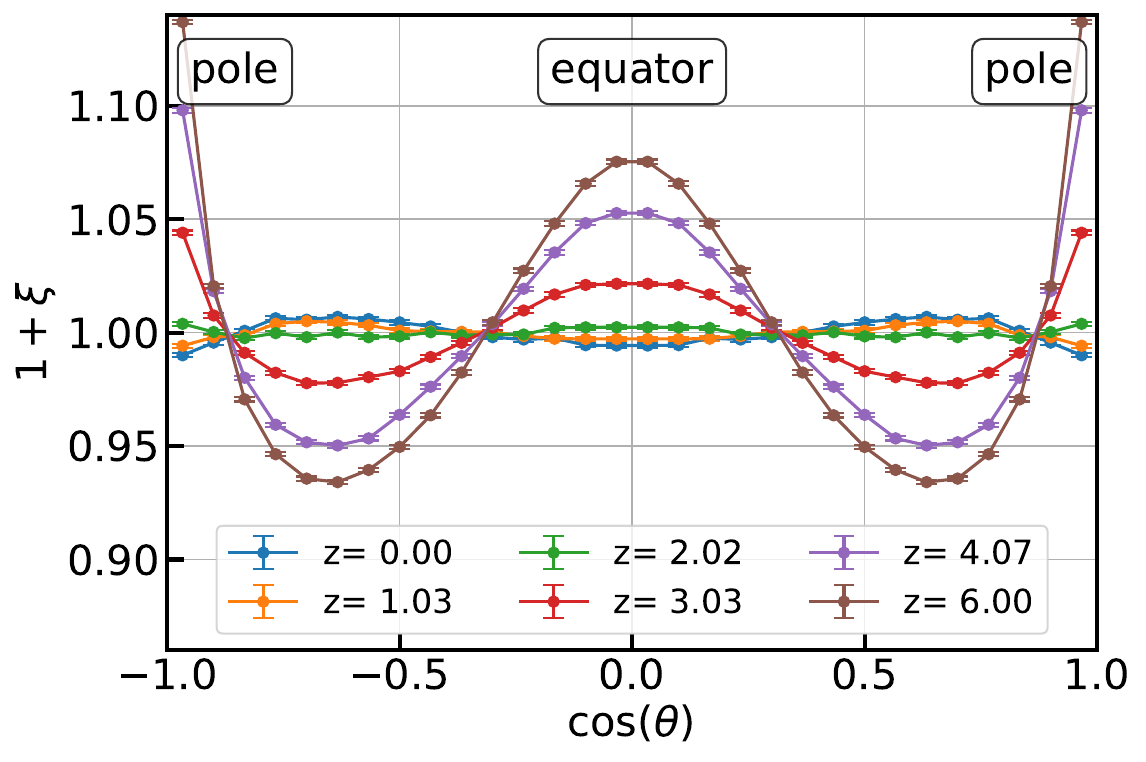}
\caption{
Similar as figure~\ref{fig:cosineexcess}, but the artifacts in minor axes are investigated.
\emph{Left panel:} The amplitude of artifacts in minor axes is same as the one in major axes.
\emph{Right panel:} At high redshifts, the minor axes have larger numerical artifacts than the major axes.
}
\label{fig:cosineminor}
\end{figure*}

\rockstar{} provides halo major axis orientations.
To keep the consistency between different results in the main text, for FoF halos we also use major axes to investigate the numerical artifacts.
However, in the initial condition, the discrete effect has a larger impact on minor axes than major axes, simply due to the shorter minor axis length by definition.
In figure~\ref{fig:elucid_minor} we show the results from the minor axes, including the alignment pattern for ELUCID FoF halos (first row),
and the alignment history of these halo particles (second row).
The minor axes do show a stronger numerical artifacts.

We also show the excess probability using minor axes of CosmicGrowth halos in figure~\ref{fig:cosineminor}.
The halo minor axes exhibit same level of alignment amplitude as the major axes.
This result is within expectation, as the principle axes are orthogonal system, as well as the three Cartesian axes of simulation box.
Thus, the alignment in one of the principle axes will also show up in another principle axis.
The halo particle alignment history is also similar when using minor axes to define the directions.
Compared to the results using major axis in figure~\ref{fig:cosineexcess}, a stronger amplitude is observed at hight redshift, as expected.

\acknowledgments
This work was supported by the National Key R\&D Program of China (No. 2023YFA1607800, 2023YFA1607801, 2023YFA1607802), the National Science Foundation of China (Grant Nos. 12273020, 12133006), the China Manned Space Project with No. CMS-CSST-2021-A03 and CMS-CSST-2025-A04, the ``111'' Project of the Ministry of Education under grant No. B20019,
and the sponsorship from Yangyang Development Fund.
The analysis is performed on the Gravity
Supercomputer at the Department of Astronomy, Shanghai Jiao Tong University.



\bibliographystyle{JHEP}
\bibliography{artifacts}

\providecommand{\href}[2]{#2}\begingroup\raggedright\begin{thebibliography}{10}

\bibitem{White83}
S.D.M.~{White}, C.S.~{Frenk} and M.~{Davis}, \emph{{Clustering in a
  neutrino-dominated universe}},
  \href{https://doi.org/10.1086/184139}{\emph{\apjl} {\bfseries 274} (1983)
  L1}.

\bibitem{White87}
S.D.M.~{White}, C.S.~{Frenk}, M.~{Davis} and G.~{Efstathiou}, \emph{{Clusters,
  Filaments, and Voids in a Universe Dominated by Cold Dark Matter}},
  \href{https://doi.org/10.1086/164990}{\emph{\apj} {\bfseries 313} (1987)
  505}.

\bibitem{Bond96}
J.R.~{Bond}, L.~{Kofman} and D.~{Pogosyan}, \emph{{How filaments of galaxies
  are woven into the cosmic web}},
  \href{https://doi.org/10.1038/380603a0}{\emph{\nat} {\bfseries 380} (1996)
  603} [\href{https://arxiv.org/abs/astro-ph/9512141}{{\ttfamily
  astro-ph/9512141}}].

\bibitem{Navarro97}
J.F.~{Navarro}, C.S.~{Frenk} and S.D.M.~{White}, \emph{{A Universal Density
  Profile from Hierarchical Clustering}},
  \href{https://doi.org/10.1086/304888}{\emph{\apj} {\bfseries 490} (1997) 493}
  [\href{https://arxiv.org/abs/astro-ph/9611107}{{\ttfamily
  astro-ph/9611107}}].

\bibitem{Bertschinger98}
E.~{Bertschinger}, \emph{{Simulations of Structure Formation in the Universe}},
  \href{https://doi.org/10.1146/annurev.astro.36.1.599}{\emph{\araa} {\bfseries
  36} (1998) 599}.

\bibitem{Springel05}
V.~{Springel}, S.D.M.~{White}, A.~{Jenkins}, C.S.~{Frenk}, N.~{Yoshida},
  L.~{Gao} et~al., \emph{{Simulations of the formation, evolution and
  clustering of galaxies and quasars}},
  \href{https://doi.org/10.1038/nature03597}{\emph{\nat} {\bfseries 435} (2005)
  629} [\href{https://arxiv.org/abs/astro-ph/0504097}{{\ttfamily
  astro-ph/0504097}}].

\bibitem{Angulo22}
R.E.~{Angulo} and O.~{Hahn}, \emph{{Large-scale dark matter simulations}},
  \href{https://doi.org/10.1007/s41115-021-00013-z}{\emph{Living Reviews in
  Computational Astrophysics} {\bfseries 8} (2022) 1}
  [\href{https://arxiv.org/abs/2112.05165}{{\ttfamily 2112.05165}}].

\bibitem{Smith24}
A.~{Smith}, C.~{Grove}, S.~{Cole}, P.~{Norberg}, P.~{Zarrouk}, S.~{Yuan}
  et~al., \emph{{Generating mock galaxy catalogues for flux-limited samples
  like the DESI Bright Galaxy Survey}},
  \href{https://doi.org/10.1093/mnras/stae1503}{\emph{\mnras} {\bfseries 532}
  (2024) 903} [\href{https://arxiv.org/abs/2312.08792}{{\ttfamily
  2312.08792}}].

\bibitem{Blake25}
C.~{Blake}, C.~{Garcia-Quintero}, S.~{Ahlen}, D.~{Bianchi}, D.~{Brooks},
  T.~{Claybaugh} et~al., \emph{{The DESI-Lensing Mock Challenge: large-scale
  cosmological analysis of 3x2-pt statistics}},
  \href{https://doi.org/10.33232/001c.131903}{\emph{The Open Journal of
  Astrophysics} {\bfseries 8} (2025) 24}
  [\href{https://arxiv.org/abs/2412.12548}{{\ttfamily 2412.12548}}].

\bibitem{Fernandez-Garcia25}
E.~{Fern{\'a}ndez-Garc{\'\i}a}, F.~{Prada}, A.~{Smith}, J.~{DeRose},
  A.J.~{Ross}, S.~{Bailey} et~al., \emph{{DESI DR2 reference mocks: clustering
  results from Uchuu-BGS and LRG}},
  \href{https://doi.org/10.48550/arXiv.2507.01593}{\emph{arXiv e-prints} (2025)
  arXiv:2507.01593} [\href{https://arxiv.org/abs/2507.01593}{{\ttfamily
  2507.01593}}].

\bibitem{Croft00}
R.A.C.~{Croft} and C.A.~{Metzler}, \emph{{Weak-Lensing Surveys and the
  Intrinsic Correlation of Galaxy Ellipticities}},
  \href{https://doi.org/10.1086/317856}{\emph{\apj} {\bfseries 545} (2000) 561}
  [\href{https://arxiv.org/abs/astro-ph/0005384}{{\ttfamily
  astro-ph/0005384}}].

\bibitem{Heavens00}
A.~{Heavens}, A.~{Refregier} and C.~{Heymans}, \emph{{Intrinsic correlation of
  galaxy shapes: implications for weak lensing measurements}},
  \href{https://doi.org/10.1046/j.1365-8711.2000.03907.x}{\emph{\mnras}
  {\bfseries 319} (2000) 649}
  [\href{https://arxiv.org/abs/astro-ph/0005269}{{\ttfamily
  astro-ph/0005269}}].

\bibitem{Xu23}
K.~{Xu}, Y.P.~{Jing}, G.-B.~{Zhao} and A.J.~{Cuesta}, \emph{{Evidence for
  baryon acoustic oscillations from galaxy-ellipticity correlations.}},
  \href{https://doi.org/10.1038/s41550-023-02035-4}{\emph{Nature Astronomy}
  {\bfseries 7} (2023) 1259}
  [\href{https://arxiv.org/abs/2306.09407}{{\ttfamily 2306.09407}}].

\bibitem{Lavaux12}
G.~{Lavaux} and B.D.~{Wandelt}, \emph{{Precision Cosmography with Stacked
  Voids}}, \href{https://doi.org/10.1088/0004-637X/754/2/109}{\emph{\apj}
  {\bfseries 754} (2012) 109}
  [\href{https://arxiv.org/abs/1110.0345}{{\ttfamily 1110.0345}}].

\bibitem{Wu25}
Z.~{Wu}, Y.~{Luo}, W.~{Wang}, X.~{Kang} and R.~{Cen}, \emph{{Cosmological
  imprints in the filament with DisPerSE}},
  \href{https://doi.org/10.1093/mnras/stae2476}{\emph{\mnras} {\bfseries 538}
  (2025) 830} [\href{https://arxiv.org/abs/2402.15712}{{\ttfamily
  2402.15712}}].

\bibitem{Efstathiou85}
G.~{Efstathiou}, M.~{Davis}, S.D.M.~{White} and C.S.~{Frenk}, \emph{{Numerical
  techniques for large cosmological N-body simulations}},
  \href{https://doi.org/10.1086/191003}{\emph{\apjs} {\bfseries 57} (1985)
  241}.

\bibitem{White94}
S.D.M.~{White}, \emph{{Formation and Evolution of Galaxies: Les Houches
  Lectures}},
  \href{https://doi.org/10.48550/arXiv.astro-ph/9410043}{\emph{arXiv e-prints}
  (1994) astro} [\href{https://arxiv.org/abs/astro-ph/9410043}{{\ttfamily
  astro-ph/9410043}}].

\bibitem{Baugh95}
C.M.~{Baugh}, E.~{Gaztanaga} and G.~{Efstathiou}, \emph{{A comparison of the
  evolution of density fields in perturbation theory and numerical simulations
  - II. Counts-in-cells analysis}},
  \href{https://doi.org/10.1093/mnras/274.4.1049}{\emph{\mnras} {\bfseries 274}
  (1995) 1049} [\href{https://arxiv.org/abs/astro-ph/9408057}{{\ttfamily
  astro-ph/9408057}}].

\bibitem{Centrella88}
J.M.~{Centrella}, J.S.~{Gallagher}, III, A.L.~{Melott} and H.A.~{Bushouse},
  \emph{{A Case Study of Large-Scale Structure in a ``Hot'' Model Universe}},
  \href{https://doi.org/10.1086/166722}{\emph{\apj} {\bfseries 333} (1988) 24}.

\bibitem{Gotz03}
M.~{G{\"o}tz} and J.~{Sommer-Larsen}, \emph{{Galaxy formation: Warm dark
  matter, missing satellites, and the angular momentum problem}},
  \href{https://doi.org/10.1023/A:1024073909753}{\emph{\apss} {\bfseries 284}
  (2003) 341} [\href{https://arxiv.org/abs/astro-ph/0210599}{{\ttfamily
  astro-ph/0210599}}].

\bibitem{Wang07}
J.~{Wang} and S.D.M.~{White}, \emph{{Discreteness effects in simulations of
  hot/warm dark matter}},
  \href{https://doi.org/10.1111/j.1365-2966.2007.12053.x}{\emph{\mnras}
  {\bfseries 380} (2007) 93}
  [\href{https://arxiv.org/abs/astro-ph/0702575}{{\ttfamily
  astro-ph/0702575}}].

\bibitem{LHuillier14}
B.~{L'Huillier}, C.~{Park} and J.~{Kim}, \emph{{Effects of the initial
  conditions on cosmological N-body simulations}},
  \href{https://doi.org/10.1016/j.newast.2014.01.007}{\emph{\na} {\bfseries 30}
  (2014) 79} [\href{https://arxiv.org/abs/1401.6180}{{\ttfamily 1401.6180}}].

\bibitem{Masaki21}
S.~{Masaki}, T.~{Nishimichi} and M.~{Takada}, \emph{{Impacts of pre-initial
  conditions on anisotropic separate universe simulations: a boosted tidal
  response in the epoch of reionization}},
  \href{https://doi.org/10.1093/mnras/staa3309}{\emph{\mnras} {\bfseries 500}
  (2021) 1018} [\href{https://arxiv.org/abs/2007.08727}{{\ttfamily
  2007.08727}}].

\bibitem{Hansen07}
S.H.~{Hansen}, O.~{Agertz}, M.~{Joyce}, J.~{Stadel}, B.~{Moore} and
  D.~{Potter}, \emph{{An Alternative to Grids and Glasses: Quaquaversal
  Pre-Initial Conditions for N-Body Simulations}},
  \href{https://doi.org/10.1086/510477}{\emph{\apj} {\bfseries 656} (2007) 631}
  [\href{https://arxiv.org/abs/astro-ph/0606148}{{\ttfamily
  astro-ph/0606148}}].

\bibitem{Joyce09}
M.~{Joyce}, B.~{Marcos} and T.~{Baertschiger}, \emph{{Towards quantitative
  control on discreteness error in the non-linear regime of cosmological N-body
  simulations}},
  \href{https://doi.org/10.1111/j.1365-2966.2008.14290.x}{\emph{\mnras}
  {\bfseries 394} (2009) 751}
  [\href{https://arxiv.org/abs/0805.1357}{{\ttfamily 0805.1357}}].

\bibitem{Liao18}
S.~{Liao}, \emph{{An alternative method to generate pre-initial conditions for
  cosmological N-body simulations}},
  \href{https://doi.org/10.1093/mnras/sty2523}{\emph{\mnras} {\bfseries 481}
  (2018) 3750} [\href{https://arxiv.org/abs/1807.03574}{{\ttfamily
  1807.03574}}].

\bibitem{Zhang21}
T.~{Zhang}, S.~{Liao}, M.~{Li} and J.~{Zhang}, \emph{{Numerical convergence of
  pre-initial conditions on dark matter halo properties}},
  \href{https://doi.org/10.1093/mnras/stab2543}{\emph{\mnras} {\bfseries 507}
  (2021) 6161} [\href{https://arxiv.org/abs/2109.02904}{{\ttfamily
  2109.02904}}].

\bibitem{Jing19}
Y.~{Jing}, \emph{{CosmicGrowth Simulations{\textemdash}Cosmological simulations
  for structure growth studies}},
  \href{https://doi.org/10.1007/s11433-018-9286-x}{\emph{Science China Physics,
  Mechanics, and Astronomy} {\bfseries 62} (2019) 19511}
  [\href{https://arxiv.org/abs/1807.06802}{{\ttfamily 1807.06802}}].

\bibitem{wang16}
H.~{Wang}, H.J.~{Mo}, X.~{Yang}, Y.~{Zhang}, J.~{Shi}, Y.P.~{Jing} et~al.,
  \emph{{ELUCID - Exploring the Local Universe with ReConstructed Initial
  Density Field III: Constrained Simulation in the SDSS Volume}},
  \href{https://doi.org/10.3847/0004-637X/831/2/164}{\emph{\apj} {\bfseries
  831} (2016) 164} [\href{https://arxiv.org/abs/1608.01763}{{\ttfamily
  1608.01763}}].

\bibitem{Ishiyama21}
T.~{Ishiyama}, F.~{Prada}, A.A.~{Klypin}, M.~{Sinha}, R.B.~{Metcalf},
  E.~{Jullo} et~al., \emph{{The Uchuu simulations: Data Release 1 and dark
  matter halo concentrations}},
  \href{https://doi.org/10.1093/mnras/stab1755}{\emph{\mnras} {\bfseries 506}
  (2021) 4210} [\href{https://arxiv.org/abs/2007.14720}{{\ttfamily
  2007.14720}}].

\bibitem{Klypin16}
A.~{Klypin}, G.~{Yepes}, S.~{Gottl{\"o}ber}, F.~{Prada} and S.~{He{\ss}},
  \emph{{MultiDark simulations: the story of dark matter halo concentrations
  and density profiles}},
  \href{https://doi.org/10.1093/mnras/stw248}{\emph{\mnras} {\bfseries 457}
  (2016) 4340} [\href{https://arxiv.org/abs/1411.4001}{{\ttfamily 1411.4001}}].

\bibitem{Chen25}
Z.~{Chen}, Y.~{Yu}, J.~{Han} and Y.~{Jing}, \emph{{CSST cosmological emulator
  I: Matter power spectrum emulation with one percent accuracy to k = 10h
  Mpc$^{‑1}$}},
  \href{https://doi.org/10.1007/s11433-025-2671-0}{\emph{Science China Physics,
  Mechanics, and Astronomy} {\bfseries 68} (2025) 289512}
  [\href{https://arxiv.org/abs/2502.11160}{{\ttfamily 2502.11160}}].

\bibitem{Chen25a}
Z.~{Chen} and Y.~{Yu}, \emph{{CSST Cosmological Emulator II: Generalized
  Accurate Halo Mass Function Emulation}},
  \href{https://doi.org/10.48550/arXiv.2506.09688}{\emph{arXiv e-prints} (2025)
  arXiv:2506.09688} [\href{https://arxiv.org/abs/2506.09688}{{\ttfamily
  2506.09688}}].

\bibitem{Zhou25}
S.~{Zhou}, Z.~{Chen} and Y.~{Yu}, \emph{{CSST Cosmological Emulator III: Hybrid
  Lagrangian Bias Expansion Emulation of Galaxy Clustering}},
  \href{https://doi.org/10.48550/arXiv.2506.04671}{\emph{arXiv e-prints} (2025)
  arXiv:2506.04671} [\href{https://arxiv.org/abs/2506.04671}{{\ttfamily
  2506.04671}}].

\bibitem{Bailin05}
J.~{Bailin} and M.~{Steinmetz}, \emph{{Internal and External Alignment of the
  Shapes and Angular Momenta of {\ensuremath{\Lambda}}CDM Halos}},
  \href{https://doi.org/10.1086/430397}{\emph{\apj} {\bfseries 627} (2005) 647}
  [\href{https://arxiv.org/abs/astro-ph/0408163}{{\ttfamily
  astro-ph/0408163}}].

\bibitem{Codis15}
S.~{Codis}, R.~{Gavazzi}, Y.~{Dubois}, C.~{Pichon}, K.~{Benabed},
  V.~{Desjacques} et~al., \emph{{Intrinsic alignment of simulated galaxies in
  the cosmic web: implications for weak lensing surveys}},
  \href{https://doi.org/10.1093/mnras/stv231}{\emph{\mnras} {\bfseries 448}
  (2015) 3391} [\href{https://arxiv.org/abs/1406.4668}{{\ttfamily 1406.4668}}].

\bibitem{Codis18}
S.~{Codis}, A.~{Jindal}, N.E.~{Chisari}, D.~{Vibert}, Y.~{Dubois}, C.~{Pichon}
  et~al., \emph{{Galaxy orientation with the cosmic web across cosmic time}},
  \href{https://doi.org/10.1093/mnras/sty2567}{\emph{\mnras} {\bfseries 481}
  (2018) 4753} [\href{https://arxiv.org/abs/1809.06212}{{\ttfamily
  1809.06212}}].

\bibitem{Wang25}
H.-d.~{Wang}, P.~{Wang}, M.~{Bao}, Y.~{Chen}, X.-x.~{Tang}, Y.~{Zhang} et~al.,
  \emph{{Observed Antiparallel Correlation between Spiral Galaxy and Cosmic
  Filament Spins}},
  \href{https://doi.org/10.3847/2041-8213/ade6fe}{\emph{\apjl} {\bfseries 987}
  (2025) L30} [\href{https://arxiv.org/abs/2506.22794}{{\ttfamily
  2506.22794}}].

\bibitem{Sousbie11}
T.~{Sousbie}, \emph{{The persistent cosmic web and its filamentary structure -
  I. Theory and implementation}},
  \href{https://doi.org/10.1111/j.1365-2966.2011.18394.x}{\emph{\mnras}
  {\bfseries 414} (2011) 350}
  [\href{https://arxiv.org/abs/1009.4015}{{\ttfamily 1009.4015}}].

\bibitem{Sousbie11a}
T.~{Sousbie}, C.~{Pichon} and H.~{Kawahara}, \emph{{The persistent cosmic web
  and its filamentary structure - II. Illustrations}},
  \href{https://doi.org/10.1111/j.1365-2966.2011.18395.x}{\emph{\mnras}
  {\bfseries 414} (2011) 384}
  [\href{https://arxiv.org/abs/1009.4014}{{\ttfamily 1009.4014}}].

\bibitem{Wang24}
W.~{Wang}, P.~{Wang}, H.~{Guo}, X.~{Kang}, N.I.~{Libeskind},
  D.~{Gal{\'a}rraga-Espinosa} et~al., \emph{{The boundary of cosmic
  filaments}}, \href{https://doi.org/10.1093/mnras/stae1801}{\emph{\mnras}
  {\bfseries 532} (2024) 4604}
  [\href{https://arxiv.org/abs/2402.11678}{{\ttfamily 2402.11678}}].

\bibitem{Yang25}
Q.-R.~{Yang}, W.~{Zhu}, G.~{Yu}, J.-F.~{Mo}, Y.~{Zheng} and L.-L.~{Feng},
  \emph{{On the width and profiles of cosmic filaments}},
  \href{https://doi.org/10.48550/arXiv.2507.02476}{\emph{arXiv e-prints} (2025)
  arXiv:2507.02476} [\href{https://arxiv.org/abs/2507.02476}{{\ttfamily
  2507.02476}}].

\bibitem{Yu25}
G.~{Yu}, W.~{Zhu}, Q.-R.~{Yang}, J.-F.~{Mo}, T.-C.~{Luan} and L.-L.~{Feng},
  \emph{{Impact of the Cosmic Web on the Properties of Galaxies in IllustrisTNG
  Simulations}}, \href{https://doi.org/10.3847/1538-4357/adc80f}{\emph{\apj}
  {\bfseries 986} (2025) 193}
  [\href{https://arxiv.org/abs/2504.01245}{{\ttfamily 2504.01245}}].

\bibitem{Tang25}
X.-x.~{Tang}, P.~{Wang}, W.~{Wang}, M.-J.~{Sheng}, H.-R.~{Yu} and H.~{Xu},
  \emph{{Cosmic Filament Spin. I. A Comparative Study in Observation}},
  \href{https://doi.org/10.3847/1538-4357/adbbd7}{\emph{\apj} {\bfseries 982}
  (2025) 197} [\href{https://arxiv.org/abs/2503.10841}{{\ttfamily
  2503.10841}}].

\bibitem{Li25}
J.~{Li}, Y.~{Zheng} and W.~{Zhu}, \emph{{Tracing Missing Baryons in the Cosmic
  Filaments with tSZ and CMB-Lensing Stacking}},
  \href{https://doi.org/10.48550/arXiv.2507.08561}{\emph{arXiv e-prints} (2025)
  arXiv:2507.08561} [\href{https://arxiv.org/abs/2507.08561}{{\ttfamily
  2507.08561}}].

\bibitem{Zhang25}
T.~{Zhang}, X.~{Liu}, Z.~{Li}, C.~{Wei}, G.~{Li}, Y.~{Luo} et~al.,
  \emph{{Modeling the impacts of galaxy intrinsic alignments on weak lensing
  peak statistics}},
  \href{https://doi.org/10.48550/arXiv.2507.09232}{\emph{arXiv e-prints} (2025)
  arXiv:2507.09232} [\href{https://arxiv.org/abs/2507.09232}{{\ttfamily
  2507.09232}}].

\end{thebibliography}\endgroup

\end{document}